\documentclass[sigconf]{acmart}

\usepackage{multirow}     
\usepackage{tabularx}     
\usepackage{subcaption}   
\usepackage{xcolor}       
\usepackage{array}        
\usepackage{booktabs}     
\usepackage{longtable}    
\usepackage{makecell}     
\usepackage{graphicx}     
\usepackage{colortbl}     
\usepackage{soul}         
\usepackage{enumitem}
\AtBeginDocument{%
  }

\begin{document}

\title{CAT-ID\texorpdfstring{$^2$}{2}: Category-Tree Integrated Document Identifier Learning for Generative Retrieval In E-commerce}

\author{Xiaoyu Liu}
\authornote{Equal contribution.}
\email{liuxiaoyv@buaa.edu.cn}
\affiliation{%
  \department{Institute of Artificial Intelligence} 
  \institution{Beihang University}
  \city{Beijing}
  \country{China}
}

\author{Fuwei Zhang}
\authornotemark[1]
\email{zhangfuwei@buaa.edu.cn}
\affiliation{%
  \department{Institute of Artificial Intelligence} 
  \institution{Beihang University}
  \city{Beijing}
  \country{China}
}

\author{Yiqing Wu}
\email{wuyiqing20s@ict.ac.cn}
\affiliation{%
  \department{Institute of Computing Technology} 
  \institution{Chinese Academy of Science}
  \city{Beijing}
  \country{China}
}

\author{Xinyu Jia}
\email{jiaxinyu04@meituan.com}
\affiliation{%
  \institution{Meituan}
  \city{Beijing}
  \country{China}
}

\author{Zenghua Xia}
\email{xiazenghua@meituan.com}
\affiliation{%
  \institution{Meituan}
  \city{Beijing}
  \country{China}
}

\author{Fuzhen Zhuang}
\authornote{Corresponding authors: Fuzhen Zhuang and Zhao Zhang} 
\email{zhuangfuzhen@buaa.edu.cn}
\affiliation{%
  \department{Institute of Artificial Intelligence}
  \institution{Beihang University}
  \city{Beijing}
  \country{China}
}
\authornote{Fuzhen Zhuang and Zhao Zhang are also at State Key Laboratory of Complex \& Critical Software Environment, Beijing, China.}

\author{Zhao Zhang}
\authornotemark[2] 
\authornotemark[3]
\email{zhangzhao.cs.ai@gmail.com}
\affiliation{%
  \institution{School of Computer Science and Engineering, Beihang University}
  \city{Beijing}
  \country{China}
}

\author{Fei Jiang}
\email{jiangfei05@meituan.com}
\affiliation{%
  \institution{Meituan}
  \city{Beijing}
  \country{China}
}

\author{Wei Lin}
\email{linwei31@meituan.com}
\affiliation{%
  \institution{Meituan}
  \city{Beijing}
  \country{China}
}

\renewcommand{\shortauthors}{Trovato et al.}

\begin{abstract}
Generative retrieval~(GR) has gained significant attention as an effective paradigm that integrates the capabilities of large language models (LLMs). It generally consists of two stages: constructing discrete semantic identifiers~(IDs) for documents and retrieving documents by autoregressively generating ID tokens.
The core challenge in GR is how to construct document IDs (DocIDS) with strong representational power. Good IDs should exhibit two key properties: similar documents should have more similar IDs, and each document should maintain a distinct and unique ID.
However, most existing methods ignore native category information, which is common and critical in E-commerce. 
Therefore, we propose a novel ID learning method, \textbf{CA}tegory-\textbf{T}ree \textbf{I}ntegrated \textbf{D}ocument \textbf{ID}entifier (CAT-ID$^2$), incorporating prior category information into the semantic IDs.
CAT-ID$^2$ includes three key modules: a Hierarchical Class Constraint Loss to integrate category information layer by layer during quantization, a Cluster Scale Constraint Loss for uniform ID token distribution, and a Dispersion Loss to improve the distinction of reconstructed documents. These components enable CAT-ID$^2$ to generate IDs that make similar documents more alike while preserving the uniqueness of different documents' representations.
Extensive offline and online experiments confirm the effectiveness of our method, with online A/B tests showing a \(0.33\%\) increase in average orders per thousand users for ambiguous intent queries and \(0.24\%\) for long-tail queries. The source code is available at \url{https://github.com/lxbdtt/CAT-ID2}.
\end{abstract}



\maketitle

\section{Introduction}
The evolution of E-commerce has positioned search systems as a pivotal tool for product discovery. In industrial settings, these systems typically adhere to a multi-stage paradigm as "\textit{semantic understanding - retrieval - ranking}". Among these components, semantic understanding often hinges on query rewriting, which standardizes vague, misspelled, or synonym-filled user queries to improve retrieval performance. The quality of this foundational step is paramount, as subsequent stages rely heavily on these rewritten queries to function effectively. 

\begin{figure}[t]
	  \centering
      \includegraphics[width=0.8\linewidth, trim=6 5 0 0, clip]{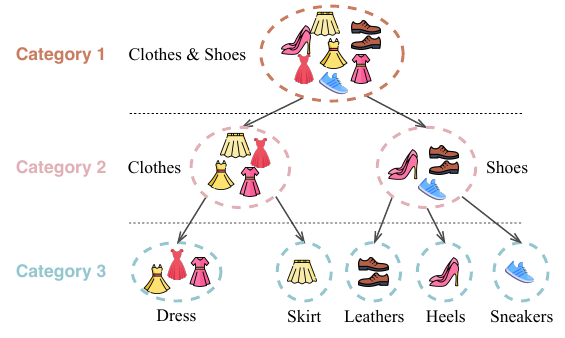}
      \caption{An example of Category-Tree in E-commerce.}
      
      \Description{A diagram illustrating the hierarchical structure of a category tree in an e-commerce system. }
       \vspace{-10pt}
      \label{fig_intro}
\end{figure}
Traditional query rewriting methods, which often rely on rule-based approaches or statistical models, struggle to effectively handle long-tail queries due to their limited ability to generalize. In contrast, LLM-based query rewriting methods~\cite{query_rewriting, liu2024query} leverage the extensive world knowledge embedded in LLMs, significantly improving rewriting performance, particularly for long-tail queries. For instance, ambiguous queries like "light meals" are more likely to be expanded into meaningful alternatives (e.g., salad, fruit, whole-grain bread) with LLMs. However, despite these advancements, the decoupled nature of this approach inevitably leads to information loss. By separating semantic understanding from retrieval, potential contextual or intent nuances in the original query may be overlooked, limiting the overall effectiveness of the search system.

The emergence of Generative Retrieval (GR) offers a groundbreaking solution to these challenges by integrating the traditionally decoupled steps of understanding and retrieval into a unified process. Unlike conventional query rewriting methods, which may lead to information loss due to their segmented nature, GR directly retrieves DocIDs based on input queries, fully leveraging the semantic capabilities of LLMs. This approach eliminates the intermediate query rewriting step~\cite{dense_1, dense_2}, thereby minimizing potential information loss and addressing the limitations of traditional systems—particularly for ambiguous or complex queries—by tightly coupling query understanding with retrieval.



GR typically operates in two stages: semantic ID indexing and autoregressive generative retrieval. The first stage discretizes the continuous semantic embeddings of documents into token ID sequences, while the second stage autoregressively generates semantic ID tokens. The effectiveness of GR heavily depends on the semantic ID indexing stage, where the discretization process acts as a form of information quantization. High-quality semantic IDs enhance LLM memory and enable accurate retrieval, while poorly constructed IDs degrade performance. An effective semantic ID construction method must satisfy three key properties: \textbf{\textit{(1) similar documents should have similar IDs; (2) dissimilar documents should have distinct IDs; and (3) semantic IDs of documents should be unique}}.

Despite significant efforts~\cite{letter,cost,zhang2025hiergr,zhang2025multi} to improve semantic representation, the use of hierarchical category information—crucial in E-commerce—remains underexplored. Most methods treat category labels as plain text during embedding, limiting their influence on semantic representations. However, hierarchical categories (as shown in Figure~\ref{fig_intro}) encode domain-specific knowledge, reflecting expert insights: documents within the same category are naturally more similar than those across categories. Yet, embedding models often fail to fully leverage this structure. Methods like RQ-VAE~\cite{rqvae} attempt to capture hierarchy via unsupervised clustering but lack the reliability of expert-defined labels. Other approaches~\cite{hi-gen} impose rigid constraints by enforcing intra-category consistency, but this comes at the cost of ignoring global semantic relationships.


In contrast, integrating category information as a soft constraint offers a balanced solution. By treating category labels as guiding signals rather than rigid boundaries, this approach preserves local category-specific information and global semantic coherence, enhancing semantic representations and retrieval performance.
To address these limitations, we propose \textbf{CA}tegory-\textbf{T}ree \textbf{I}ntegrated \textbf{D}ocument \textbf{ID}entifier (CAT-ID\(^2\)), a novel ID construction method that incorporates hierarchical category tree into the indexing process. Specifically, we use Residual Quantization Variation Encoder (RQ-VAE)~\cite{rqvae, rq-image} to quantize semantic vectors and introduce three key losses: the Hierarchical Class Constraint Loss, which ensures intra-category compactness and inter-category separation; the Cluster Scale Constraint Loss, which prevents encoding collapse; and the Dispersion Loss, which promotes diversity among semantic IDs. These components enable CAT-ID\(^2\) to generate IDs aligned with the semantic structure of e-commerce data, improving downstream LLM learning and retrieval performance.
Our contributions are summarized as follows:
\begin{itemize}[leftmargin=*, topsep=0pt, itemsep=1pt, parsep=0pt]
    \item We propose a novel ID construction method integrating hierarchical category tree labels, an essential but previously underutilized feature in E-commerce, into the document indexing process. 
    \item We propose a novel combination of loss functions—Hierarchical Class Constraint Loss, Cluster Scale Constraint Loss, and Dispersion Loss—to construct IDs with key semantic properties.
    \item We conduct extensive offline and online experiments to demonstrate the effectiveness of CAT-ID$^2$. 
\end{itemize}
\section{Related Work}
\noindent\textbf{Sparse and Dense Retrieval.}
Typical sparse retrieval methods~\cite{bm25, tf-idf} calculate term-document matching scores based on sparse text representation but often face challenges like lexical mismatches. Dense Retrieval~\cite{dpr, ance} encodes queries and documents into dense vectors and performs retrieval using Max Inner Product Search. Therefore, DR primarily focuses on improving encoding quality through approaches such as hard negative mining~\cite{hard_mining}, knowledge distillation~\cite{kd}, query enhancement with external knowledge~\cite{zhang2022mind}, and utilizing more advanced encoders~\cite{reimers2019sentence}.

\noindent\textbf{Generative Retrieval.} 
GR constructs IDs using lexical information such as titles~\cite{grgr} and urls~\cite{ultron} or hierarchical semantic structures~\cite{dsi, lc-rec}. The former leverages the generation capabilities of LLMs but struggles to scale to large datasets, while the latter generates hierarchical IDs yet often overlooks hierarchical category labels. Hi-Gen~\cite{hi-gen} is the first to explicitly integrate category labels into ID generation but uses a rigid prefix-based approach that lacks flexibility and fails to capture global semantic relationships. Additionally, some studies~\cite{seal, nci} propose constrained decoding to ensure generated IDs remain within the valid ID sequences.

\begin{figure*}[t]
	  \centering
      \includegraphics[width=0.92\linewidth]{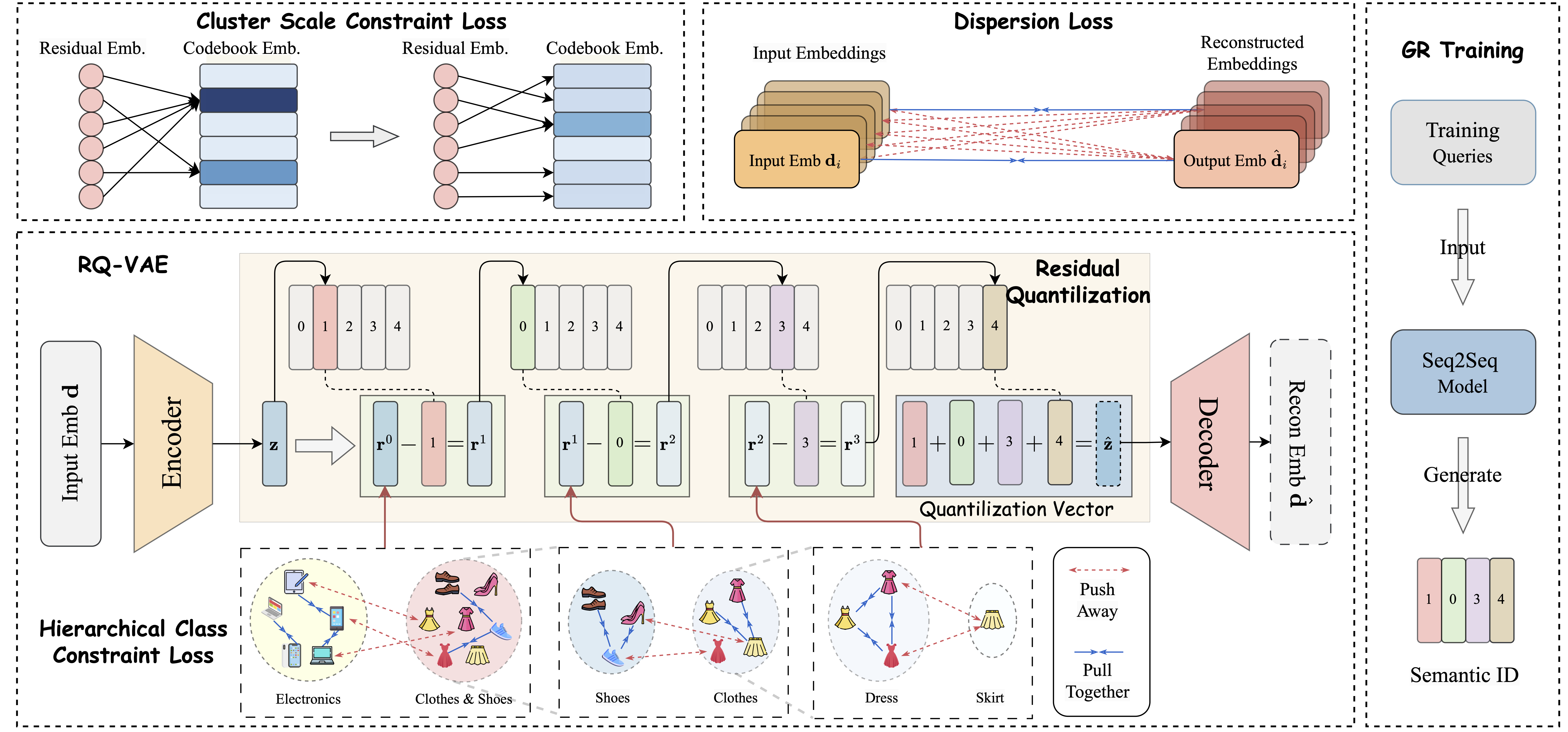}
      \caption{Overall framework of CAT-ID$^2$. It comprises two stages: DocID Learning and Generative Model Training. In the DocID Learning, Hierarchical Class Constraint Loss $\mathcal{L}_\text{HCC}$, Cluster Scale Constraint Loss $\mathcal{L}_\text{CSC}$, and Dispersion Loss $\mathcal{L}_\text{Dis}$ are introduced alongside the original residual quantinize loss $\mathcal{L}_\text{rq}$. HCCL integrates prior category information, CSCL ensures uniform codebook utilization to prevent collapse, and Dispersion Loss enforces distinct semantic IDs for different documents.} 
      \label{fig_method}
     \vspace{-8pt}
\end{figure*}

\section{Methodology}
We briefly introduce the process of RQ-VAE. As shown in Figure~\ref{fig_method}, given a document set \(\mathcal{D}=\{(d_1, C_1),\dots,(d_n, C_n)\}\), where \(d_i\) represents the document-specific information and \(C_i=\{c_{i}^1, \dots, c_{i}^H\}\) denotes its hierarchical category labels with \(H\) as the category depth, the input data is first encoded. A sequence-to-sequence model (e.g., BERT~\cite{bert} or T5~\cite{t5}) encodes the concatenation of \(d_i\) and \(C_i\) into a text embedding \(\mathbf{d}\), which is then mapped to a latent representation \(\mathbf{z}=E (\mathbf{d})\) using a DNN Encoder \(E(\cdot)\). The quantization process involves \(L\) quantizer layers, each equipped with a codebook \(\mathcal{C}^l = \{\mathbf{e}^l_k\}_{k=1}^{K}\), where \(K\) is the size of the codebook. At layer \(l\), the residual vector \(\mathbf{r}^l\) is compared with codebook vectors to compute a probability distribution \(\mathbf{p}^l=\text{Softmax}(-\|\mathbf{r}^l-\mathbf{e}^l_k\|_2)\), 
and the quantization result is \(c^l={\arg\max}_k \mathbf{p}^l\). After \(L\) layers of quantization, a coarse-to-fine ID sequence \((c^0, \dots, c^{L-1})\) is generated.

The quantized representation \(\hat{\mathbf{z}} = \sum_{l=0}^{L-1} \mathbf{e}_{c_l}\) is used to approximate the original latent representation \(\mathbf{z}\). The training objective is defined as:
\begin{equation}
    \mathcal{L}_{\text{recon}} = \|\mathbf{d} - D(\hat{\mathbf{z}})\|_2^2,
\end{equation}
\begin{equation}
    \mathcal{L}_{\text{rq}}\! =\!\sum_{l}^{L}\! \left( \|\text{sg}[\mathbf{r}_l]\! -\! \mathbf{e}_{c_l}^l\|_2^2 \!+\! \eta \|\mathbf{r}_l\! -\! \text{sg}[\mathbf{e}_{c_l}^l]\|_2^2 \right),
\end{equation}
\begin{equation}
    \mathcal{L}_{\text{RQ-VAE}}  = \mathcal{L}_{\text{recon}} + \mathcal{L}_{\text{rq}},
\end{equation}
where \(D(\cdot)\) is the Decoder and \(\text{sg}[\cdot]\) denotes the stop-gradient operation.

\subsection{Hierarchical Class Constraint Loss}
In E-commerce, hierarchical category information is both prevalent and crucial. Documents within the same category naturally exhibit higher similarity, making the hierarchical structure highly advantageous for semantic ID representation. As mentioned above, effective DocIDs ensure similar documents have similar IDs, while dissimilar ones are distinct. However, existing methods~\cite{dsi, hi-gen} that rely on hard partitioning and intra-category clustering often sacrifice global semantic information. To address this, we propose the Hierarchical Class Constraint Loss (HCCL), which incorporates the InfoNCE~\cite{infonce} loss into the quantizer layers for contrastive learning. The loss function is defined as:

\begin{equation}
    \mathcal{L}^l_{\text{HCC}} = -\frac{1}{|B|}\sum_{\mathbf{r}^l_a\in B}  \log 
    \frac{\exp(\langle \mathbf{r}^l_{a}, \mathbf{r}^l_{p}\rangle / \tau)}
    {\sum\limits_{\mathbf{r}^l_n\in B} \exp(\langle 
 \mathbf{r}^l_a, \mathbf{r}^l_n\rangle / \tau)},
 \label{HCCL}
\end{equation}
where \( \mathbf{r}^l_a \), \( \mathbf{r}^l_p \), and \( \mathbf{r}^l_n \) represent the anchor, positive, and negative samples. \( B \) is the data batch, \( \tau \) is the temperature coefficient, and \( \langle \cdot, \cdot \rangle \) refers to the cosine similarity. Each quantizer layer corresponds to a specific category depth, leveraging both global semantic information and category labels.

In our approach, documents within the same category are treated as positive examples, while those from different categories are treated as negative examples. To further enhance inter-category separability, starting from the second quantizer layer, we adopt a hard negative mining strategy. Specifically, we select negative examples that belong to the same category in the previous layer but fall into different subcategories in the current layer. This strategy effectively encourages semantically similar documents to cluster closer together while pushing apart dissimilar ones. It is worth noting that \textbf{the maximum depth \( H \) of the category tree must be smaller than the maximum depth \( L \) of the RQ-VAE}. This ensures that the model retains sufficient learning capacity and avoids collapsing into a simple category-tree structure.

\subsection{Cluster Scale Constraint Loss}
\label{ps:collapse}
While HCCL effectively enforces hierarchical constraints, it may encounter a collapse issue during optimization. For instance, in the first layer, if the number of item categories \(|C^1|\) is smaller than the number of codebook entries \(K\), samples from each category might exclusively occupy a single codebook entry. While this satisfies the constraints of HCCL, it leads to inefficient utilization of the codebook. To address this, we propose the Cluster Scale Constraint Loss (CSCL), defined as: 

\begin{equation}
    \mathcal{L}_{\text{CSC}}^l = \text{KL}(\frac{\sum\limits_{i}^{|B|}\mathbf{p}^l_i}{|B|}||\frac{1}{K}\cdot\mathbf{1})+\text{KL}(\frac{1}{K}\cdot\mathbf{1}||\frac{\sum\limits_{i}^{|B|}\mathbf{p}^l_i}{|B|}),
\end{equation}
where \(\text{KL}(\cdot)\) denotes the KL-divergence, \(\mathbf{p}^l_i\) represents the probability distribution over codebook entries for the \(i\)-th document, and \(\mathbf{1}\) is a vector of ones. The bidirectional KL-divergence in CSCL serves two purposes: the first term discourages the overuse of certain codebook entries, while the second penalizes unused entries. CSCL penalizes imbalanced assignments by encouraging the average distribution of samples across codebook entries to approach uniformity. Consequently, it significantly improves the utilization efficiency of the codebook.

\subsection{Dispersion Loss}
Different documents should have distinct semantic IDs. To enhance the distinctiveness of generated IDs, we introduce the Dispersion Loss. The key idea is to encourage reconstructed embeddings \(\hat{\mathbf{z}}\) to be as distinct as possible. This translates to maximizing the dissimilarity of \(\hat{\mathbf{d}} = D(\hat{\mathbf{z}})\) across all data points. 
Therefore, we adopt the InfoNCE loss to achieve this, defined as:

\begin{equation}
    \mathcal{L}_{\text{dif}} = -\frac{1}{|B|}\sum_{\hat{\mathbf{z}}_i\in B}  \log 
    \frac{\exp(\langle \hat{\mathbf{z}}_i, \hat{\mathbf{z}}_i\rangle / \tau)}
    {\sum\limits_{\hat{\mathbf{z}}_j\in B} \exp(\langle 
 \hat{\mathbf{z}}_i, \hat{\mathbf{z}}_j\rangle / \tau)},
\end{equation}
where \(\hat{\mathbf{z}}_i\) denotes the reconstructed embedding of a sample, and \(\hat{\mathbf{z}}_j\) represents other reconstructed embeddings in the same batch \(B\). This loss pushes all reconstructed embeddings apart, enhancing distinctiveness.
When combined with the reconstruction loss \(\mathcal{L}_{\text{recon}}= \|\mathbf{d} - D(\hat{\mathbf{z}})\|_2^2\), which ensures \(\hat{\mathbf{d}}\) closely approximates the original embedding \(\mathbf{d}\), the overall objective balances two goals: making \(\hat{\mathbf{d}}_i\) similar to its original embedding \(\mathbf{d}_i\)  while being dissimilar to embeddings of other samples \(\mathbf{d}_j\) (\(j \neq i\)). We redefine this objective as the Dispersion Loss~(DisL), expressed as:

\begin{equation}
    \mathcal{L}_{\text{Dis}}= -\frac{1}{|B|}\sum_{\hat{\mathbf{d}}_i\in B}  \log 
    \frac{\exp(\langle \hat{\mathbf{d}}_i, {\mathbf{d}}_i\rangle / \tau)}
    {\sum\limits_{{\mathbf{d}}_j\in B} \exp(\langle 
 \hat{\mathbf{d}}_i, {\mathbf{d}}_j\rangle / \tau)}.
\end{equation}

This loss can directly replace \(\mathcal{L}_{\text{recon}}\). Interestingly, we note that the form of this loss is similar to a contrastive loss used in CoST~\cite{cost}. However, it is important to emphasize that our motivation is fundamentally different. The loss in CoST aims to capture essential neighborhood relationships for effective item modeling in recommender systems, whereas our DisL is specifically designed to ensure the distinctiveness of reconstructed embeddings for ID generation.

\subsection{Training \& Inference}
\textbf{ID Tokenization}. 
The model is trained by combining \(\mathcal{L}_{\text{rq}}\) with the three constraint losses introduced earlier. The overall objective function is defined as:

\begin{equation}
    \mathcal{L}_{\text{ID}}=\mathcal{L}_{\text{rq}}+ 
    \alpha\mathcal{L}_{\text{Dis}}+
    \beta\sum_l^L\mathcal{L}_{\text{HCC}}^l+ 
    \gamma\sum_l^L\mathcal{L}_{\text{CSC}}^l, 
\end{equation}
where \(\alpha\), \(\beta\), and \(\gamma\) are hyperparameters balancing different losses. For DocIDs that remain colliding after being generated by the RQ-VAE, we adopt the Sinkhorn algorithm~\cite{cuturi2013sinkhorn} as a post-processing step to reassign unique DocIDs.

\noindent\textbf{Sequence Modeling}. 
Following TIGER~\cite{tiger}, we fine-tune the LLM using the generated ID sequences. New tokens are introduced into the LLM, and the model is optimized using the following objective:

\begin{equation}
    \mathcal{L}_\text{Seq} = -\sum_{t=1}^{T} \log P(y_t | y_{<t}, q),
\end{equation}
where \(y_t\) denotes the token at time step \(t\), \(y_{<t}\) represents all preceding tokens, and \(q\) is the input query. This objective enables the model to learn the conditional probability distribution over ID sequences of a given query. During inference, Beam Search is used to generate multiple candidate document IDs for a given query \(q\).
\section{Experiments}
In this section, we analyze the effectiveness of our proposed model, CAT-ID\(^2\), and address the following research questions. \textbf{RQ1}: How does CAT-ID\(^2\) perform compared to other baselines? \textbf{RQ2}: What is the contribution of each module in CAT-ID\(^2\) to the overall performance? \textbf{RQ3}: What is the quality of DocID generated by CAT-ID\(^2\)? \textbf{RQ4}: How do different hyperparameters affect CAT-ID\(^2\)? \textbf{RQ5}: What is the training efficiency of CAT-ID\(^2\)? \textbf{RQ6}: How does CAT-ID\(^2\) perform in real-world deployment scenarios?

\begin{table}  
    \centering
    \caption{Statistics of processed ESCI datasets. }
    \vspace{-6pt}
    \setlength\tabcolsep{5.5pt}
    \begin{tabular}{l|c|c@{\hspace{4pt}}c|c@{\hspace{4pt}}c}
        \toprule
        \multirow{3}{*}{Dataset} & \multirow{3}{*}{Docs } & \multicolumn{2}{c|}{Train} & \multicolumn{2}{c}{Test} \\
        \cmidrule(lr){3-4} \cmidrule(lr){5-6}
        & & Queries & Q-D Pairs & Queries & Q-D Pairs \\
        \midrule
        ESCI-us & 386,392 & 20,001 & 341,460 & 5,800 & 29,351 \\
        ESCI-es & 131,935 & 5,350 & 121,297 & 1,612 & 12,176 \\
        ESCI-jp & 152,845 & 6,458 & 139,508 & 1,832 & 14,353 \\
        \bottomrule
    \end{tabular}
    \label{tab:statics}
    \vspace{-8pt}
\end{table}
\subsection{Settings}
\noindent\textbf{Datasets.} 
We use the ESCI\footnote{\url{https://github.com/amazon-science/esci-data}}~\cite{esci}, a publicly available multilingual large-scale E-commerce search dataset, including queries and product information in \textbf{English~(us)}, \textbf{Spanish~(es)}, and \textbf{Japanese~(jp)}. Each query is associated with up to 40 potentially relevant products labeled by relevance levels: Exact~(E), Substitute~(S), Complement~(C), and Irrelevant~(I), and products include multi-level category information with depths ranging from 2 to 7. 
For preprocessing, we remove products whose category depth is less than 3. For those with a category depth greater than 3, we truncate the depth to 3.
During sequence modeling, only query-document pairs with relevance levels of E and S are used for training. Additionally, all products in the test set are ensured to also appear in the training set. The dataset statistics are summarized in Table~\ref{tab:statics}. 


\begin{table*}[t] 
    \centering
\renewcommand{\arraystretch}{1.05}

    \setlength\tabcolsep{2.5pt} 
    \caption{Performance of different models on ESCI datasets (us, es, jp) in terms of Recall@5, 10, 20, 50, 100 (\%). The best results are highlighted in \textbf{bold}, while the second-best results are \underline{underlined}. * indicates the best performance among GR methods. 256 and 512 are different codebook sizes.} 
    \vspace{-6pt}
    \begin{tabular}{c|ccccc|ccccc|ccccc}
        \toprule
        \multirow{2}{*}{Model} & \multicolumn{5}{c|}{\textbf{ESCI-us} (Recall)} & \multicolumn{5}{c|}{\textbf{ESCI-es} (Recall)} & \multicolumn{5}{c}{\textbf{ESCI-jp} (Recall)} \\
        \cmidrule{2-16}
               & @5   & @10  & @20  & @50  & @100 & @5   & @10  & @20  & @50  & @100 & @5   & @10  & @20  & @50  & @100 \\
        \midrule
        BM-25  & 3.35 & 5.68 & 9.05 & 14.69 & 19.42 & 3.10 & 5.19 & 8.40 & 13.75 & 18.85 & 2.56 & 3.23 & 3.85 & 4.45 & 4.86 \\
        \midrule
        DPR    & \underline{3.84} & \textbf{6.77} & \textbf{10.88} & \textbf{18.76} & \textbf{26.34} & 2.77 & 4.64 & 7.63 & 13.73 & 19.50 & 2.86 & 4.45 & 6.45 & 10.05 & 14.10 \\
        Sen-T5 & {3.04} & 4.98 & {8.32} & {14.80} & \underline{23.47} & -- & -- & -- & -- & -- & -- & -- & -- & -- & -- \\
        MPNet&1.70&2.88&4.62&8.62&13.08&1.69&3.04&4.94&8.00&11.69&0.57&0.85&1.19&1.65&2.07\\
        \midrule
 DSI\textsubscript{naive}& 0.19 & 0.27 & 0.38 & 0.66 & 1.05 & 1.22 & 2.22 & 3.66 & 6.09 & 8.84 & 1.05 & 1.76 & 2.91 & 5.07 & 6.73 \\
 DSI\textsubscript{semantic}& 1.29 & 2.09 & 3.23 & 5.31 & 7.51 & 4.61 & 7.93 & 12.96 & 21.37 & 27.28 & 3.06 & 5.52 & 9.44 & 16.55 & 21.47 \\
        Hi-Gen & 1.82 & 2.93 & 4.40 & 7.00 & 9.60 & 4.87 & 8.31 & 13.60 & 22.00 & 28.15 & 3.44 & 6.08 & 10.75 & 17.32 & 22.96\\
        NCI    & 2.67 & 4.55 & 6.48 & 10.59 & 16.82 & 5.16 & 8.98 & 14.11 & 22.56 & 29.32 & 3.82 & 6.95 & 11.47 & 17.65 & 23.71 \\
        TIGER  & 2.86 & 4.93 & 7.86 & 13.42 & 18.59 & \underline{5.70} & \underline{9.45} & \underline{15.31} & \underline{24.14} & \underline{31.08} & \underline{4.32} & \underline{7.64} & \underline{12.40} & \underline{20.21} & \underline{25.83} \\
        \midrule
    CAT-ID$^2$(256) & 3.64 & 5.86 & 8.99 & 15.17& 21.00 & \textbf{6.09}* & \textbf{10.14}* & \textbf{15.97}* & \textbf{24.79}* & \textbf{32.10}* & \textbf{4.90}* & 7.89 & 12.39 & 20.23 & 26.61 \\
    CAT-ID$^2$(512) & \textbf{3.97}* & \underline{6.54}* & \underline{10.21}* & \underline{16.75}*& {23.37}* & 5.38 & 9.71 & 15.50 & 24.73 & 31.60 & 4.57 & \textbf{8.09}* & \textbf{13.06}* & \textbf{21.88}* & \textbf{28.14}* \\
        \bottomrule
    \end{tabular}
    \label{tab:Main}
    \vspace{-6pt}
\end{table*}

\noindent\textbf{Baselines}.
To evaluate our model, we compare it against three baseline categories: sparse, dense, and GR models. For sparse retrieval, we use BM25~\cite{bm25}. For dense retrieval, we include DPR~\cite{dpr}, Sentence-T5~\cite{sentence-t5}, and Multilingual MPNet~\cite{song2020mpnet}. For GR models, we evaluate DSI~\cite{dsi}, NCI~\cite{nci}, Hi-Gen~\cite{hi-gen} and TIGER~\cite{tiger}. 

\noindent\textbf{Evaluation}.
We use \textbf{Recall@k} for evaluation. Similar to prior works~\cite{genret, hi-gen}, we do not use NDCG because ranking performance is not the primary concern at the recall stage. \textbf{Recall@k} measures the proportion of relevant items retrieved within the top-k results, reflecting the model's capability to identify relevant candidates.

\noindent\textbf{Implementation Details.}
We reproduce the baselines on the ESCI dataset using open-source code, with all experiments utilizing the T5-base model~\cite{t5} as the backbone for GR models. For RQ-VAE, we set the codebook to \( L = 4 \) layers and trained it for 300 epochs. We use \( K = 256 \) and \( K = 512 \) entries per layer. The batch size \(|B|\) is set to 4096, and we use the AdamW optimizer. The hyperparameters are configured as \(\alpha=0.1\), \(\beta=0.0001\), and \(\gamma=1.0\). All experiments are conducted on a platform with eight A100 80GB GPUs. Since a multilingual version of Sentence-T5 is not available, we only report results on ESCI-us.

\subsection{Overall Perfomance (RQ1)}
Table~\ref{tab:Main} presents the experimental results. 
Here, we derive the following observations:
1) CAT-ID$^2$ achieves the best performance among all GR methods. In addition, our model also outperforms all DR models on the ESCI-es and ESCI-jp datasets. This demonstrates the effectiveness of the proposed method, particularly in the GR paradigm.
2) Our model substantially improves upon TIGER, which itself surpasses DSI\textsubscript{semantic}. This hierarchy highlights that high-quality identifiers are a critical factor for enhancing model performance. 
3) The optimal codebook size is dependent on the data scale. Larger datasets require bigger codebooks, while smaller datasets suffer performance loss from overly large ones.
4) Hi-Gen's performance is comparable to DSI-Semantic because it similarly relies on hierarchical K-means clustering. This approach \textbf{lacks the global semantic information captured by RQ-VAE-based methods, resulting in inferior performance} relative to TIGER and CAT-ID$^2$.

\begin{table}[t] 
    \centering
    \renewcommand{\arraystretch}{1.05}
   \caption{Ablation Study on ESCI-us. ``+'' indicates the addition of a specific module} 
    \vspace{-6pt}
    \begin{tabular}{l|>{\centering\arraybackslash}p{.95cm}%
                      >{\centering\arraybackslash}p{.95cm}%
                      >{\centering\arraybackslash}p{.95cm}}
    \toprule
    \multicolumn{1}{c|}{Model} & R@10 & R@50 & R@100\\
    \midrule
    (0)  \quad TIGER & 4.93 & 13.42& 18.60\\
    \midrule
    (1) \, (0)+HCCL & 4.76 & 12.81 & 17.99\\
    (2) \, (0)+CSCL & 4.83 & 12.65 & 17.84\\
    (3) \, (1)+CSCL & \underline{5.35} & \underline{14.69} & \underline{19.05}\\
    (4) \, (0)+DisL & 5.10 & 13.58 & 18.83\\
    (5) \, (1)+DisL & 5.17 & 13.76 & 18.95\\
    \midrule
    (6) \, \, CAT-ID$^2$ & \textbf{5.86} & \textbf{15.17} & \textbf{21.00}\\
    \bottomrule
    \end{tabular}
    \vspace{-6pt}
    \label{tab:ablation}
\end{table}

\subsection{Ablation Study (RQ2)}
To investigate the roles of different modules in CAT-ID$^2$, we conduct ablation studies on ESCI-us. The experimental results of several model variants are shown in Table~\ref{tab:ablation}, where ``+'' indicates the addition of a specific module to a variant. Based on these results, we draw the following conclusions: 
1) Using HCCL or CSCL individually imposes overly strong constraints on specific model aspects, leading to degraded performance. 2) HCCL alone is prone to model collapse (as discussed in Section~\ref{ps:collapse}). However, combining it with either CSCL or DisL effectively alleviates this issue and improves performance, demonstrating the complementary nature of these loss functions. 3) The best performance is achieved when all three losses (HCCL, CSCL, and DisL) are applied together (CAT-ID$^2$). This is likely because DisL provides a strong stabilizing effect, enabling HCCL and CSCL to function as effective constraints without over-restricting the codebook distribution. This balanced interaction significantly enhances the model's performance.

As shown in Table~\ref{cate_info}, we also analyze the impact of removing explicit category information from the input (denoted as "w/o cate info"). This led to only a minor performance drop across all models, \textbf{confirming our hypothesis that the impact of category information would be limited due to the presence of other rich textual features in the embeddings}, reducing the model's reliance on this explicit feature.  In particular, our model achieves the smallest drop in recall when category information is removed, highlighting its effectiveness.

\begin{table}[t]
\centering
\renewcommand{\arraystretch}{1.05}
\caption{Ablation study about category information. "wo cate info" denotes removing category information in embedding.}
\vspace{-6pt}
\begin{tabular}{c|ccccc}
\toprule
\multirow{2}{*}{\textbf{Model}} & \multicolumn{5}{c}{\textbf{Recall}} \\ 
\cmidrule(lr){2-6}
 & @5 & @10 & @20 & @50 & @100 \\
\midrule
DSI-Semantic & 1.29 & 2.10 & 3.24 & 5.31 & 7.51 \\
 wo cate info & 1.22 & 1.93 & 3.10 & 5.15 & 7.33 \\
\midrule
TIGER & 2.86 & 4.94 & 7.86 & 13.42 & 18.60 \\
wo cate info & 2.70 & 4.85 & 7.69 & 13.27 & 18.23 \\
\midrule
\textbf{CAT-ID$^2$} & \textbf{3.64} & \textbf{5.87} & \textbf{8.99} & \textbf{15.17} & \textbf{21.00} \\
wo cate info & \underline{3.60} & \underline{5.79} & \underline{8.90} & \underline{15.13} & \underline{20.95} \\
\bottomrule
\end{tabular}
\label{cate_info}
\vspace{-6pt}
\end{table}

\begin{table}[t]
\centering
\renewcommand{\arraystretch}{1.05} 
\caption{Ablation study on category depth.}
\vspace{-6pt}
\begin{tabular}{c|cccccc}
\toprule
\multirow{2}{*}{\textbf{Model}} & \multicolumn{5}{c}{\textbf{Recall}} \\ 
\cmidrule(lr){2-6}
 & @5 & @10 & @20 & @50 & @100 \\
\midrule
TIGER & 2.86 & 4.93 & 7.86 & 13.42 & 18.59\\
Random Sampling & \underline{3.50} & \underline{5.61} & \underline{8.76} & \underline{14.78} & \underline{20.34}\\
\midrule
1 level & 3.19 & 5.33 & 8.34 & 13.99 & 19.00\\
2 level  & 3.28 & 5.46 & 8.57 & 14.24 & 19.21 \\
2\&3 level & 3.46 & 5.57 & 8.68 & 14.49 & 19.96 \\
\midrule
3 level (Ours) & \textbf{3.65} & \textbf{5.87} & \textbf{8.99} & \textbf{15.17} & \textbf{21.00} \\
\bottomrule
\end{tabular}
\vspace{-6pt}
\label{depth}
\end{table}

\begin{figure}[t]
    \centering
    \begin{subfigure}[b]{0.92\linewidth}
        \centering
        \includegraphics[width=\textwidth]{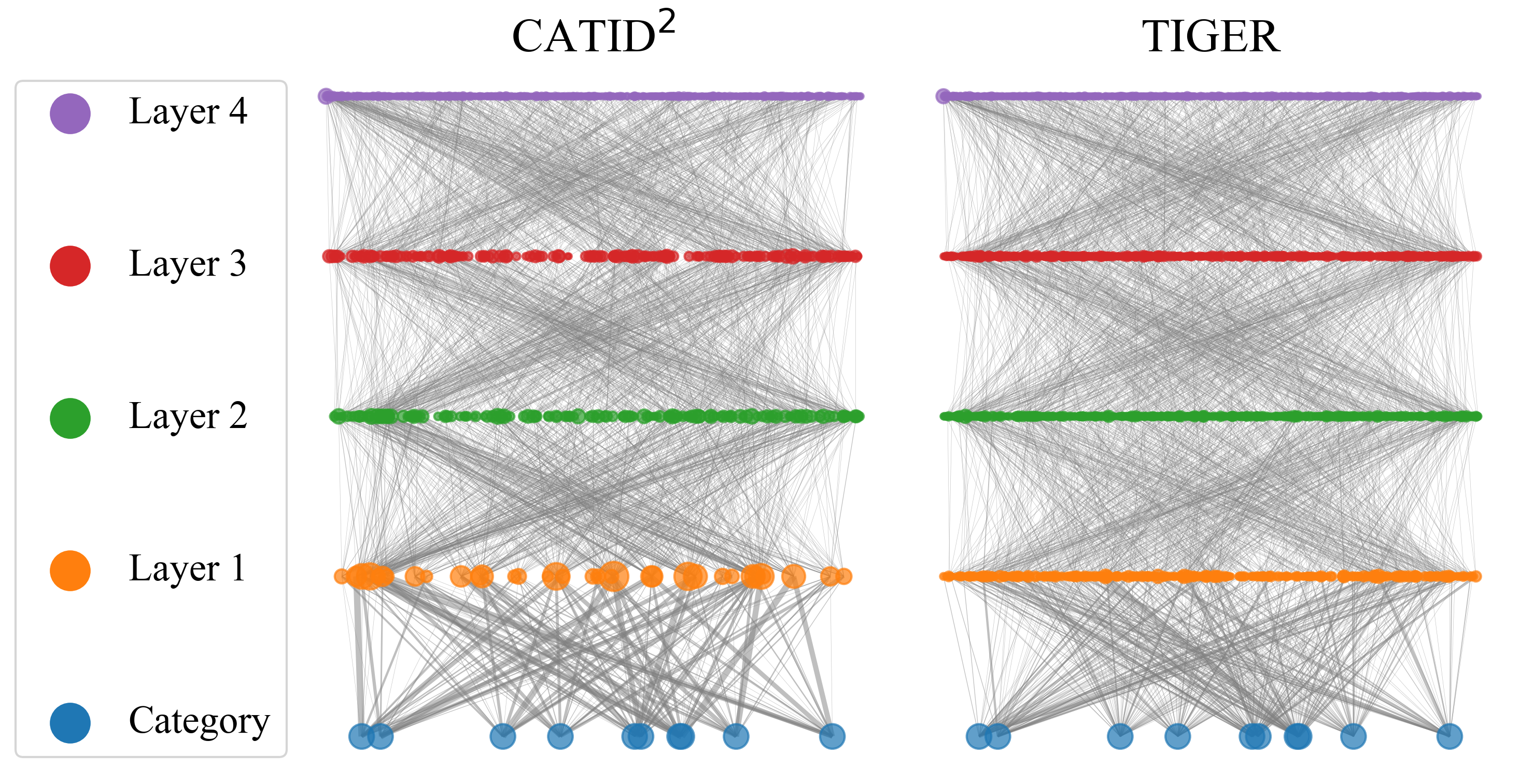}
        \caption{The first category level.}
        \label{fig:distribution_layer1}
    \end{subfigure}
    \begin{subfigure}[b]{0.92\linewidth}
        \centering
        \includegraphics[width=\textwidth]{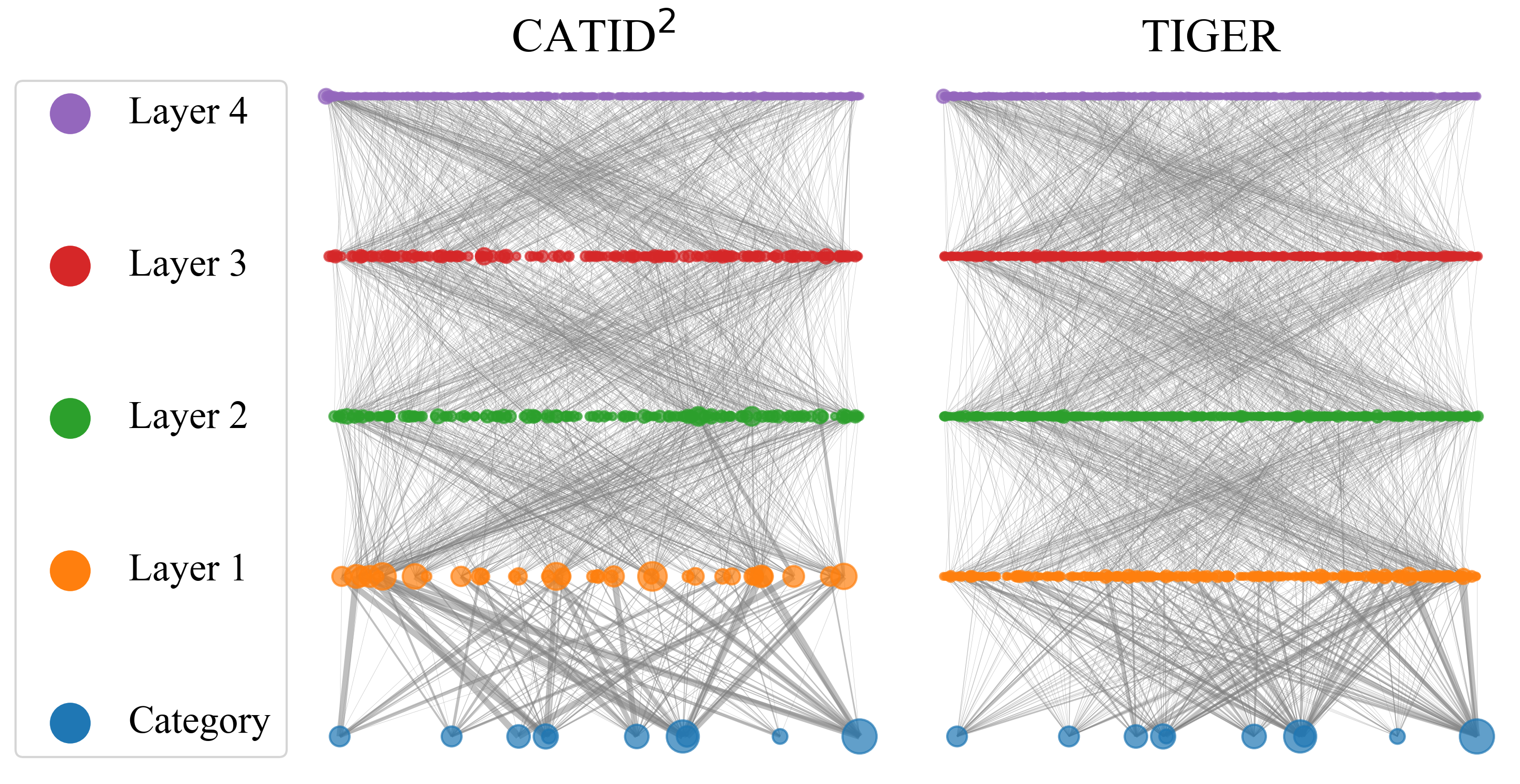}
        \caption{The second category level.}
        \label{fig:distribution_layer2}
    \end{subfigure}
    \begin{subfigure}[b]{0.92\linewidth}
        \centering
        \includegraphics[width=\textwidth]{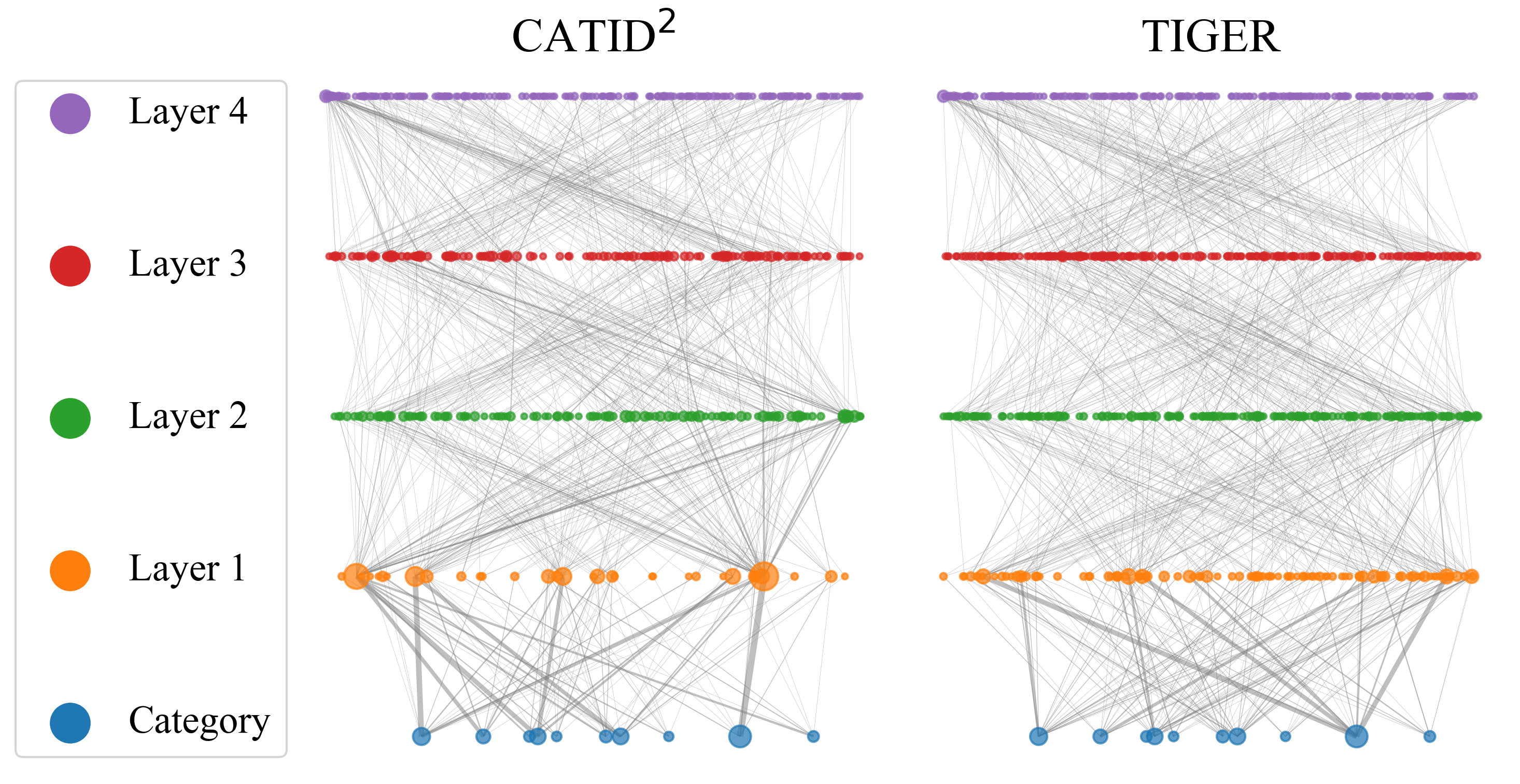}
        \caption{The third category level.}
        \label{fig:distribution_layer3}
    \end{subfigure}
    \vspace{-6pt}
    \caption{Layer distribution of IDs across the top 10 categories. Larger nodes indicate a greater number of samples within the corresponding codebook.}
    \label{fig:layer1and2and3}
\end{figure}

\begin{figure*}[t]
    \centering
     \begin{subfigure}[b]{0.27\textwidth}
        \centering
        \includegraphics[width=\textwidth]{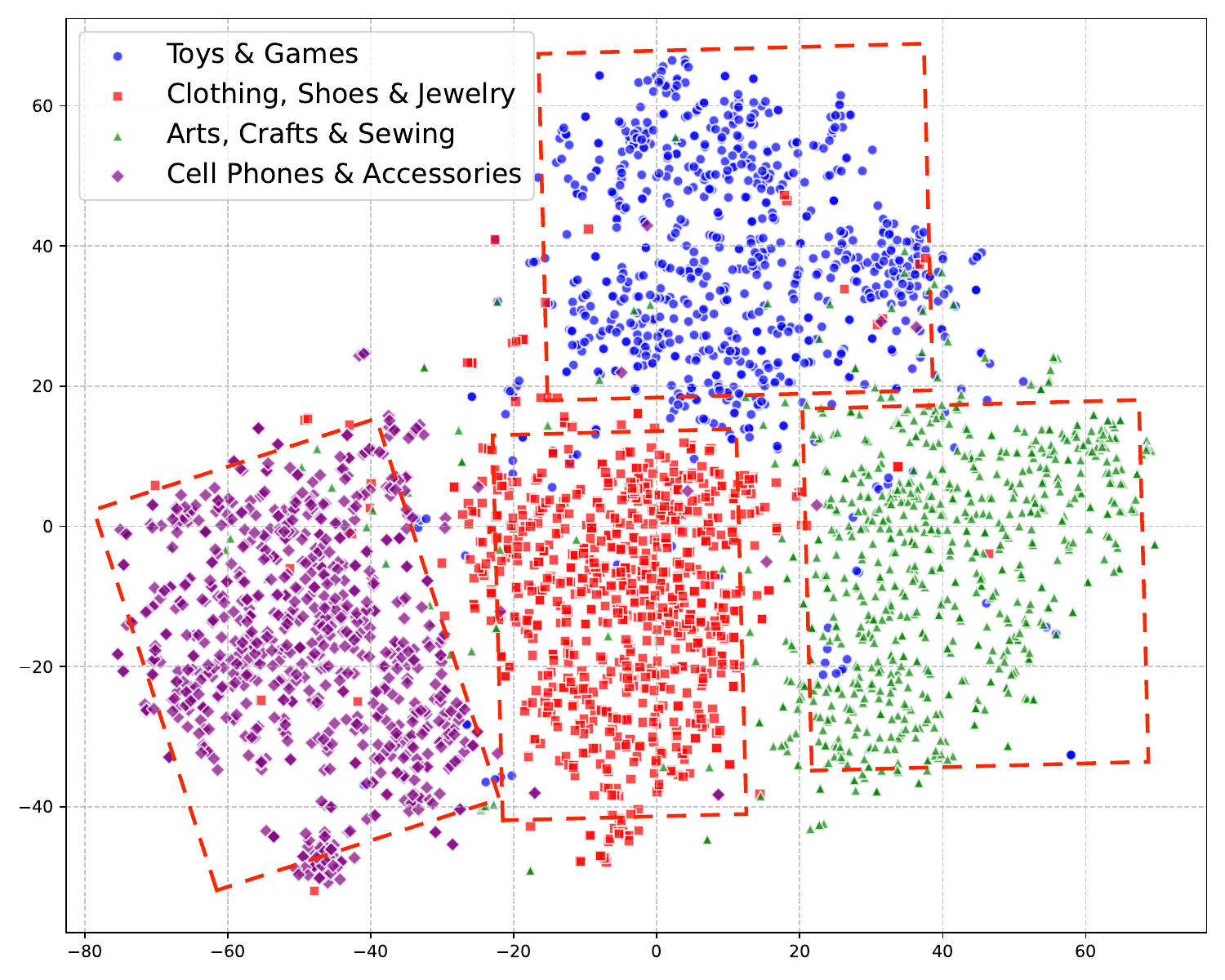}
        \caption{CAT-ID$^2$}
        \label{fig:catid2_tsne}
        \vspace{-6pt}
    \end{subfigure}
    \hspace{0.05\textwidth}
    \begin{subfigure}[b]{0.27\textwidth}
        \centering
        \includegraphics[width=\textwidth]{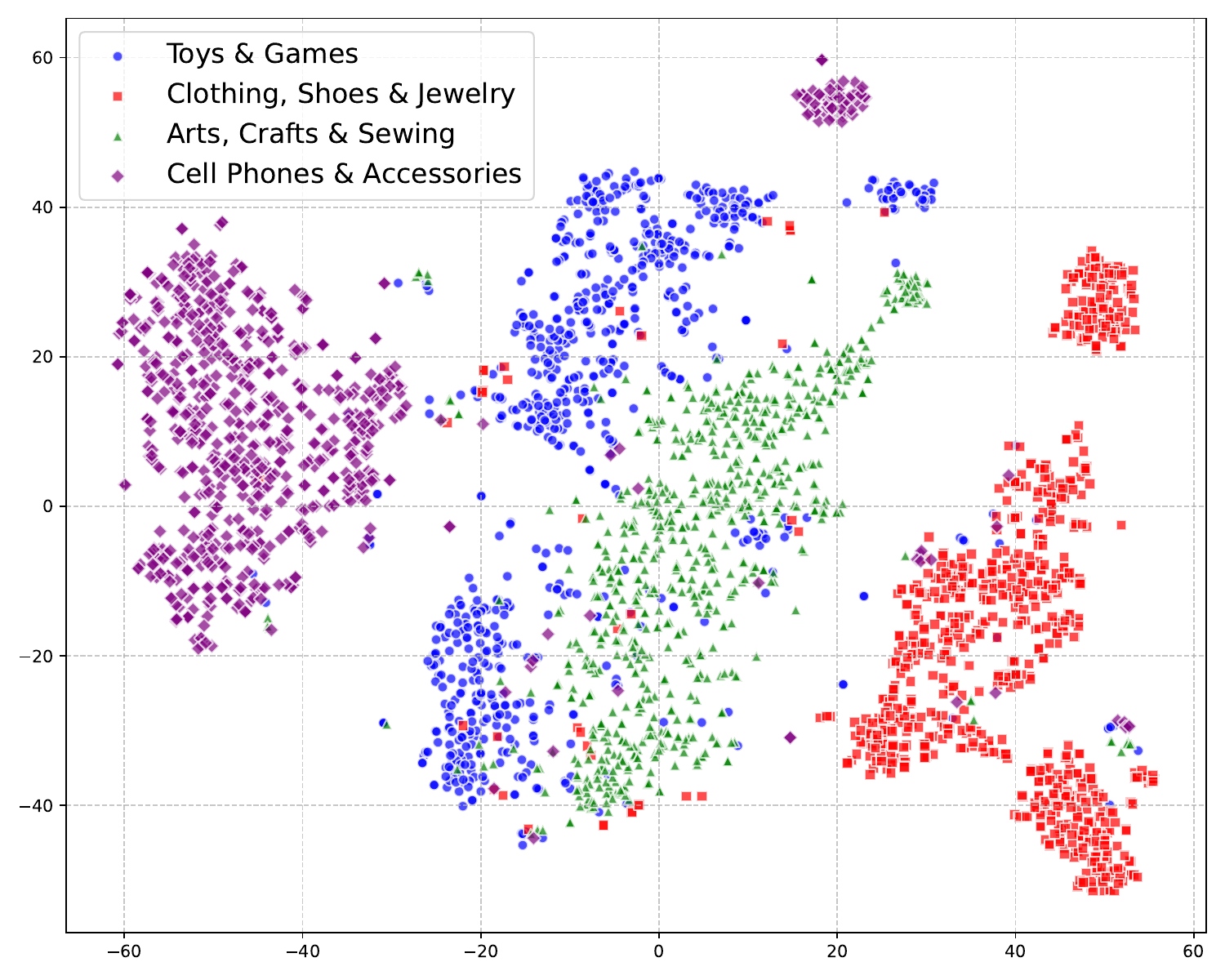}
        \caption{TIGER}
        \label{fig:tiger_tsne}
        \vspace{-6pt}
    \end{subfigure}
    \hspace{0.05\textwidth}
    \begin{subfigure}[b]{0.27\textwidth}
        \centering
        \includegraphics[width=\textwidth]{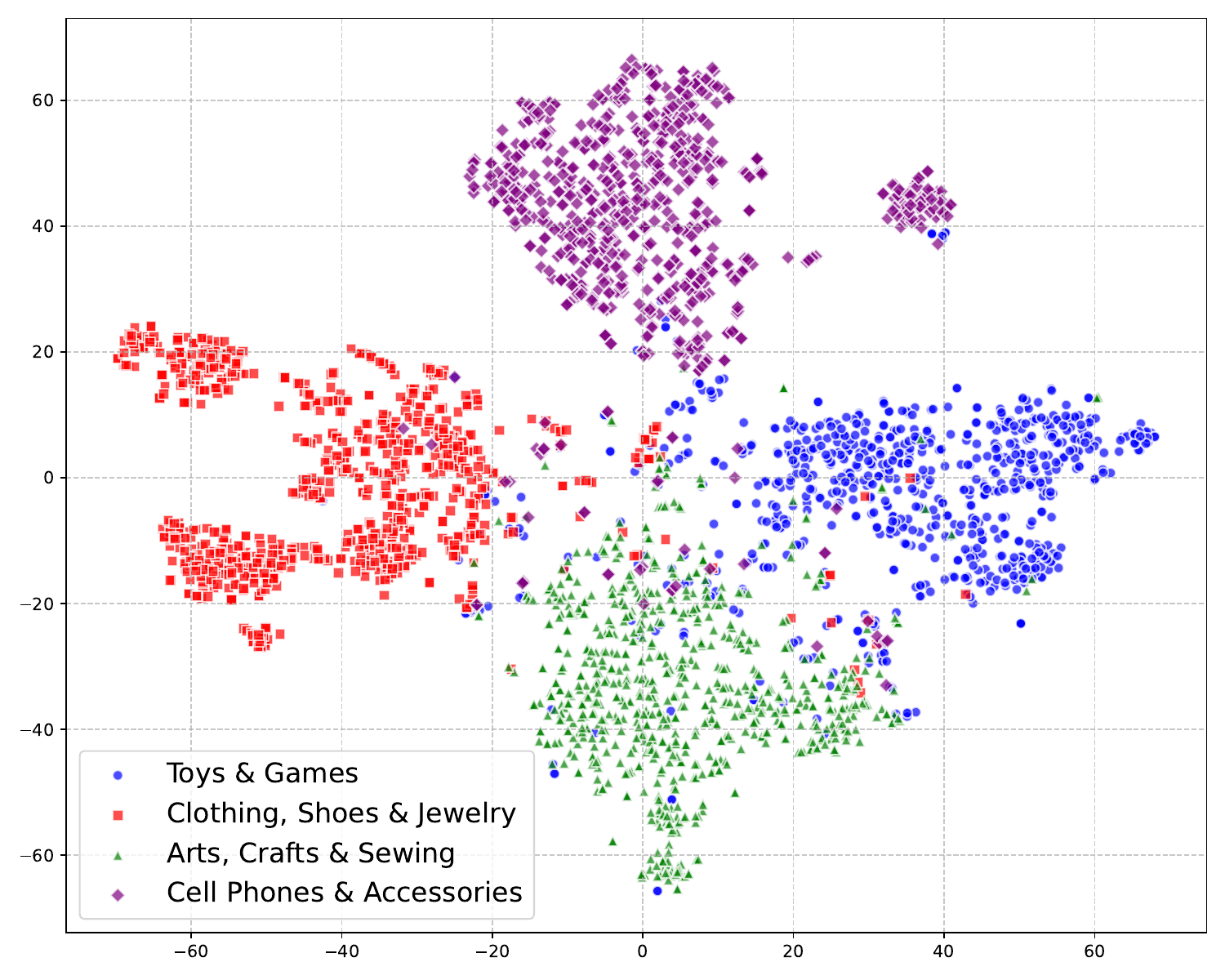}
        \caption{Only HCCL}
        \label{fig:hccl_tsne}
        \vspace{-6pt}
    \end{subfigure}
    \caption{Visualization of different documents under four categories: 1) Toy \& Games. 2) Clothing, Shoes \& Jewelry. 3) Arts, Crafts \& Sewing. 4) Cell Phones \& Accessories.}
    \vspace{-6pt}
    \label{fig:tsne}
\end{figure*}

\subsection{Analysis of Generated DocIDs (RQ3)}
To better understand the distribution across different layers, we visualize the connections among DocID layers in Figure~\ref{fig:layer1and2and3}. Blue nodes represent distinct categories. Nodes in other colors correspond to the four layers of codebooks, with each layer containing 512 codebooks. Node size reflects the number of samples per codebook, and edge thickness denotes the degree of overlap between connections. The results demonstrate that CAT-ID$^2$ effectively achieves assigning similar IDs to documents within the same category. On the other hand, the distributions in the higher layers of DocIDs resemble those produced by TIGER, displaying a more dispersed and less clustered pattern. This highlights CAT-ID$^2$'s ability to generate DocIDs that are not only semantically meaningful but also distinct, leveraging document hierarchies in the process.

We further visualize document representations encoded by the RQ-VAE DNN encoder using t-SNE~\cite{van2008visualizing} (Figure~\ref{fig:tsne}).  Within the same category, CAT-ID$^2$ produces more compact representations compared to TIGER, avoiding the issue of scattered points. However, when only HCCL is used, the intra-category distinguishability is insufficient. The introduction of CSCL and DisL effectively addresses this issue, achieving a balance where representations are relatively compact within categories without collapsing entirely. This ensures the clear distinguishability of different item IDs.

\begin{table}[t]
\centering
\renewcommand{\arraystretch}{1.05}
\caption{Impact of the number of the layers of RQ-VAE.}
\vspace{-6pt}
\begin{tabular}{c|ccccc}
\toprule
\multirow{2}{*}{\textbf{Model}} & \multicolumn{5}{c}{\textbf{Recall}} \\ 
\cmidrule(lr){2-6}
 & @5 & @10 & @20 & @50 & @100 \\
\midrule
TIGER(3 layers) & 2.88 & 4.87 & 7.44 & 12.64 & 17.38 \\
\textbf{CAT-ID$^2$(3 layers)} & {2.97} & {5.00} & {7.83} & {12.93} & {17.50} \\
\midrule
TIGER(4 layers) & 2.86 & 4.94 & 7.86 & 13.42 & 18.60 \\
\textbf{CAT-ID$^2$(4 layers)} & \textbf{3.64} & \textbf{5.87} & \textbf{8.99} & \textbf{15.17} & \textbf{21.00} \\
\midrule
TIGER(5 layers) & 3.06 & 5.01 & 7.87 & 13.40 & 17.99 \\
\textbf{CAT-ID$^2$(5 layers)} & \underline{3.13} & \underline{5.12} & \underline{8.11} & \underline{13.68} & \underline{19.03} \\
\bottomrule

\end{tabular}
\vspace{-6pt}
\label{layers}
\end{table}

\begin{figure*}[t]
    \centering
    \begin{subfigure}[b]{0.20\textwidth}
        \centering
        \includegraphics[width=\textwidth]{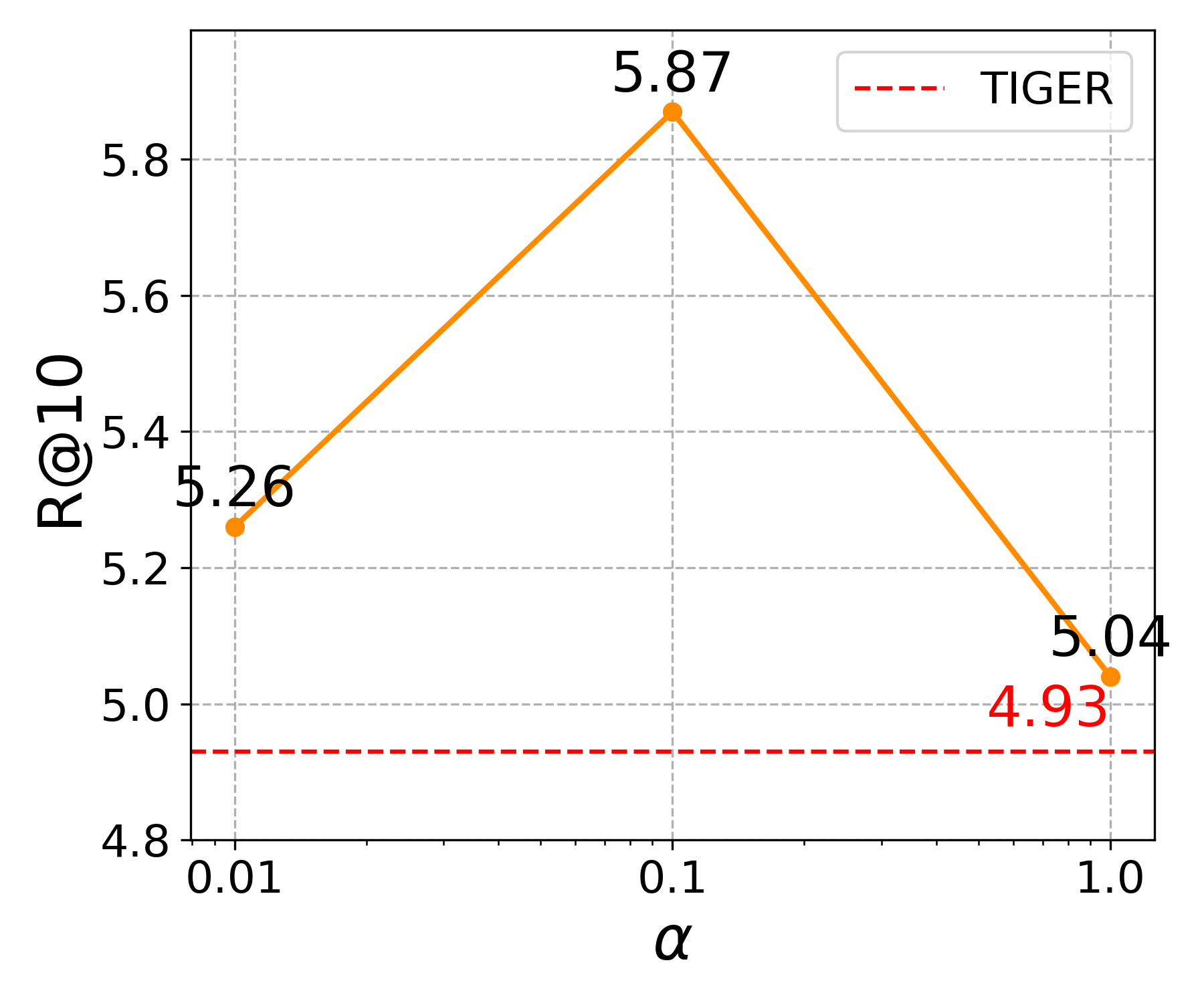}
        \caption{$R@10$ with $\alpha$}
        \label{fig:alpha_10}
    \end{subfigure}
    \hspace{0.04\textwidth}
    \begin{subfigure}[b]{0.20\textwidth}
        \centering
        \includegraphics[width=\textwidth]{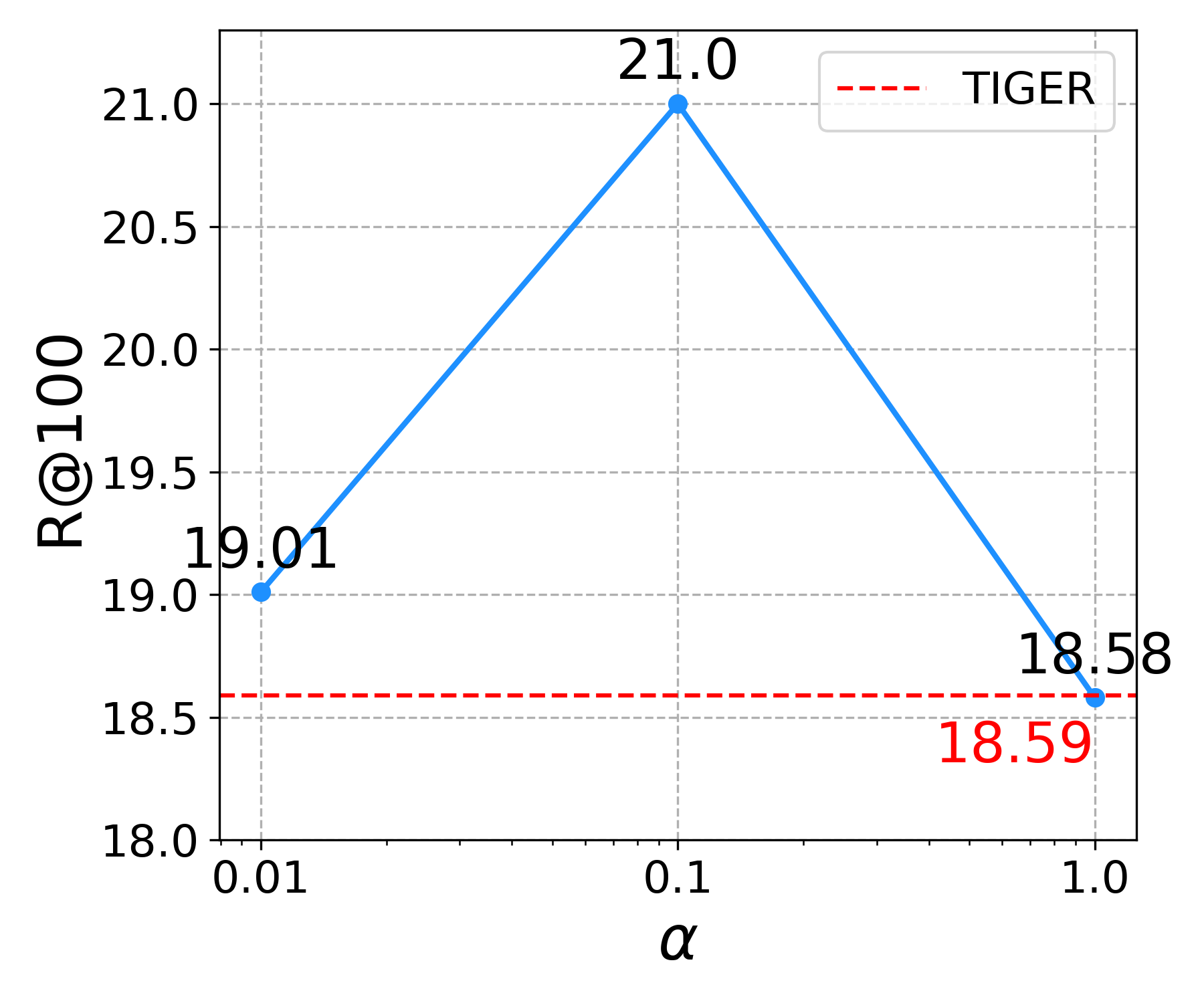}
        \caption{$R@100$ with $\alpha$}
        \label{fig:alpha_100}
    \end{subfigure}
    \hspace{0.04\textwidth}
    \begin{subfigure}[b]{0.20\textwidth}
        \centering
        \includegraphics[width=\textwidth]{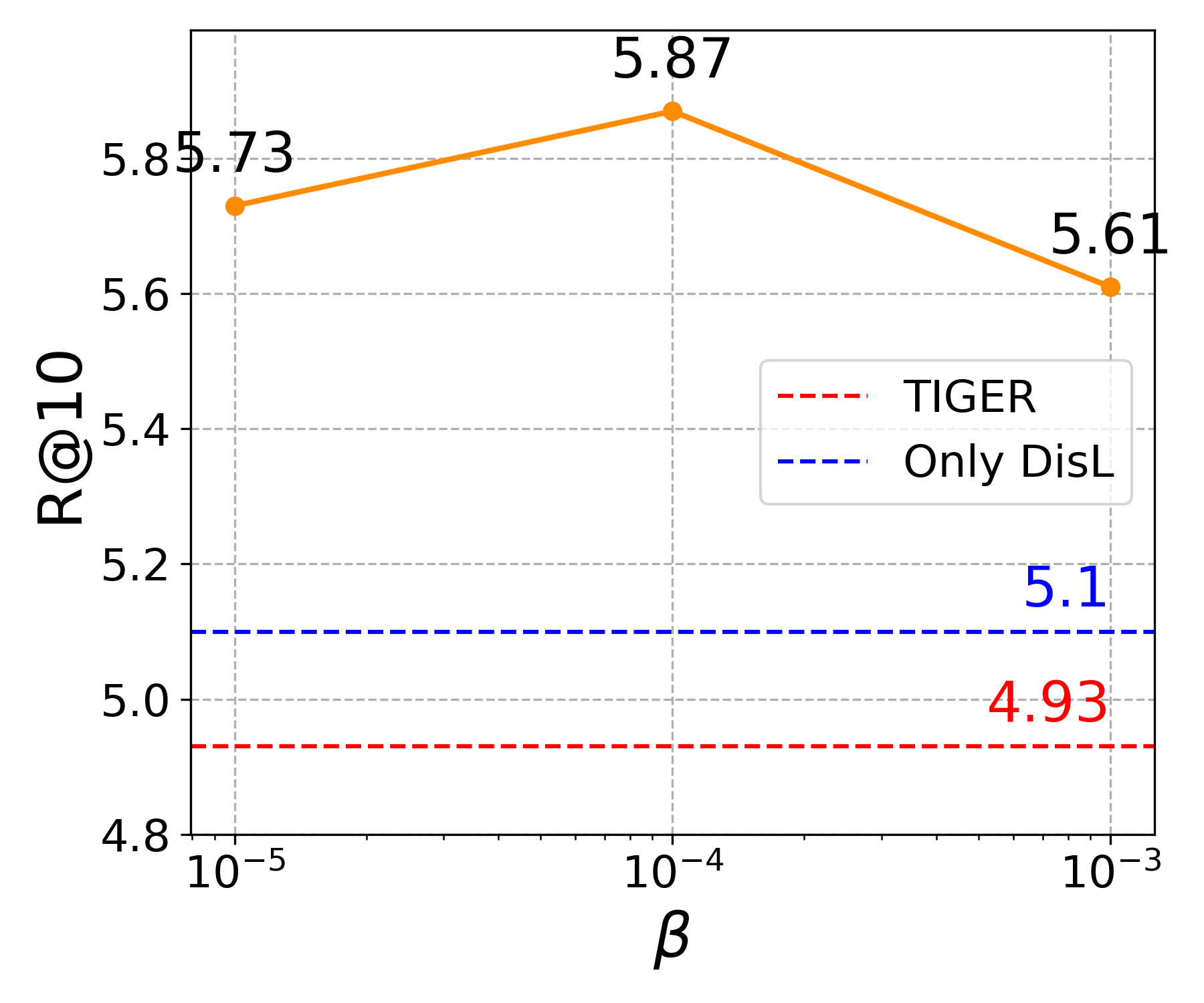}
        \caption{$R@10$ with $\beta$}
        \label{fig:beta_10}
    \end{subfigure}
    \hspace{0.04\textwidth}
    \begin{subfigure}[b]{0.20\textwidth}
        \centering
        \includegraphics[width=\textwidth]{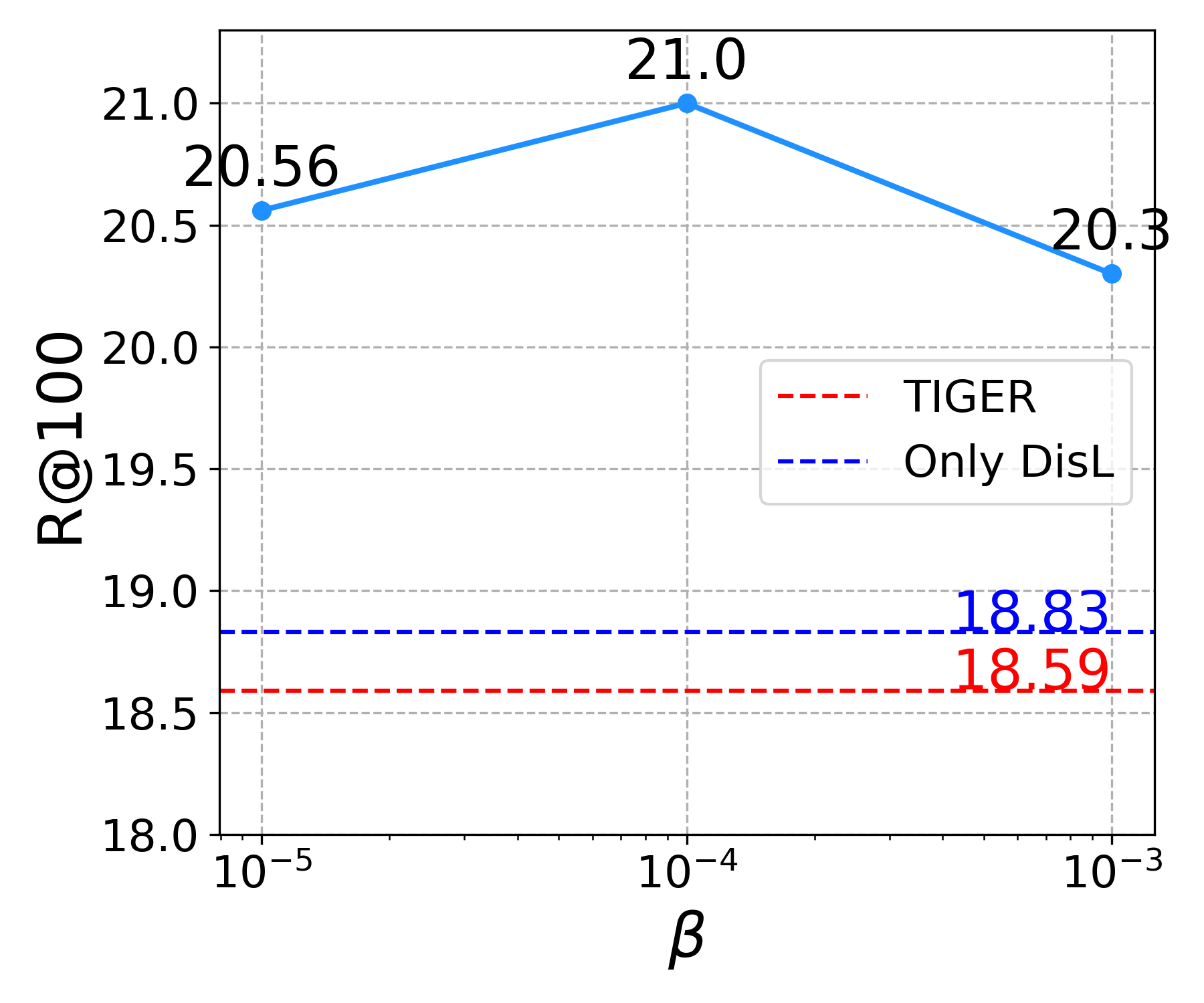}
        \caption{$R@100$ with $\beta$}
        \label{fig:beta_100}
    \end{subfigure}
    \hspace{0.04\textwidth}
    \begin{subfigure}[b]{0.20\textwidth}
        \centering
        \includegraphics[width=\textwidth]{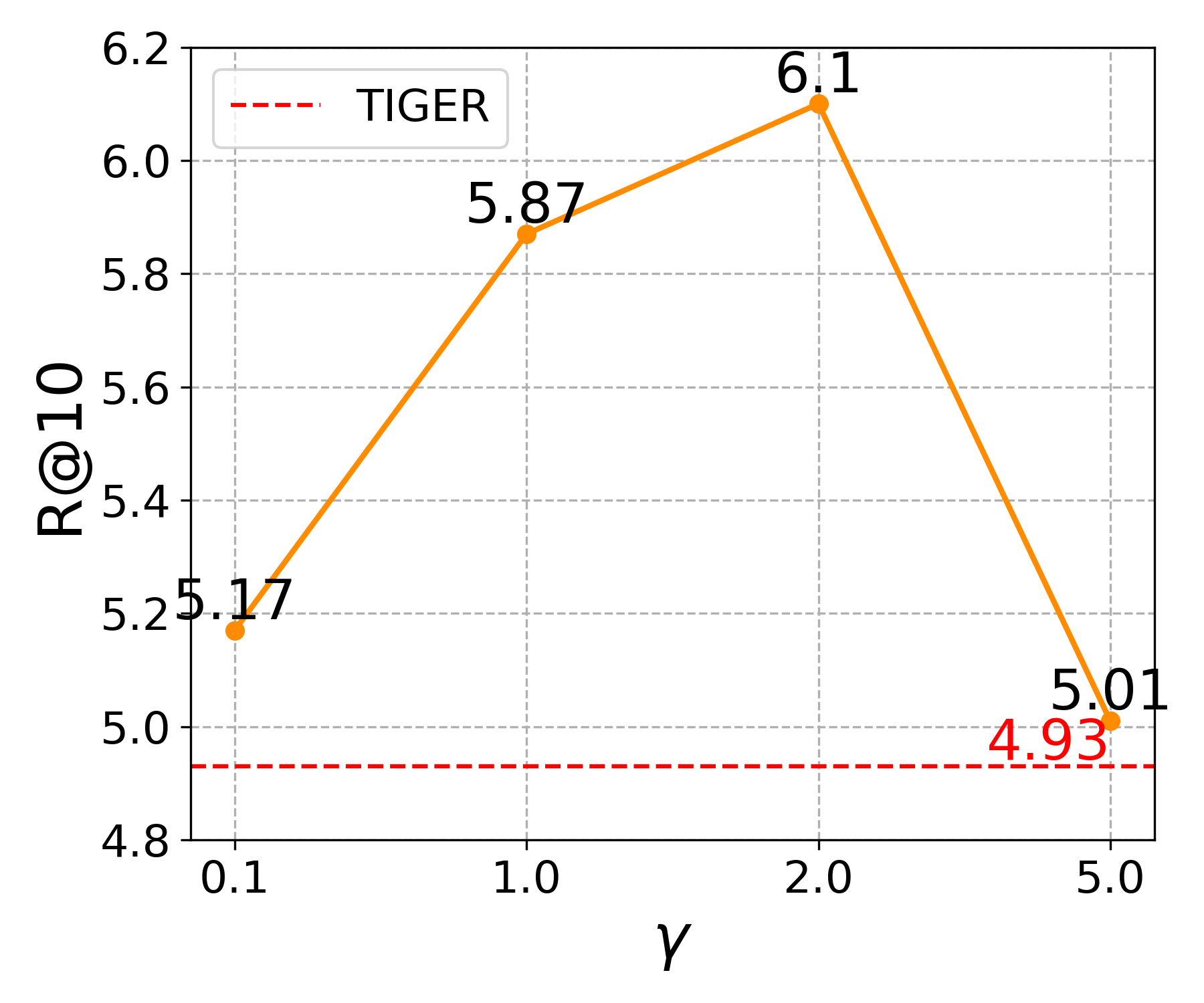}
        \caption{$R@10$ with $\gamma$}
        \label{fig:gamma_10}
        \vspace{-6pt}
    \end{subfigure}
    \hspace{0.04\textwidth}
    \begin{subfigure}[b]{0.20\textwidth}
        \centering
        \includegraphics[width=\textwidth]{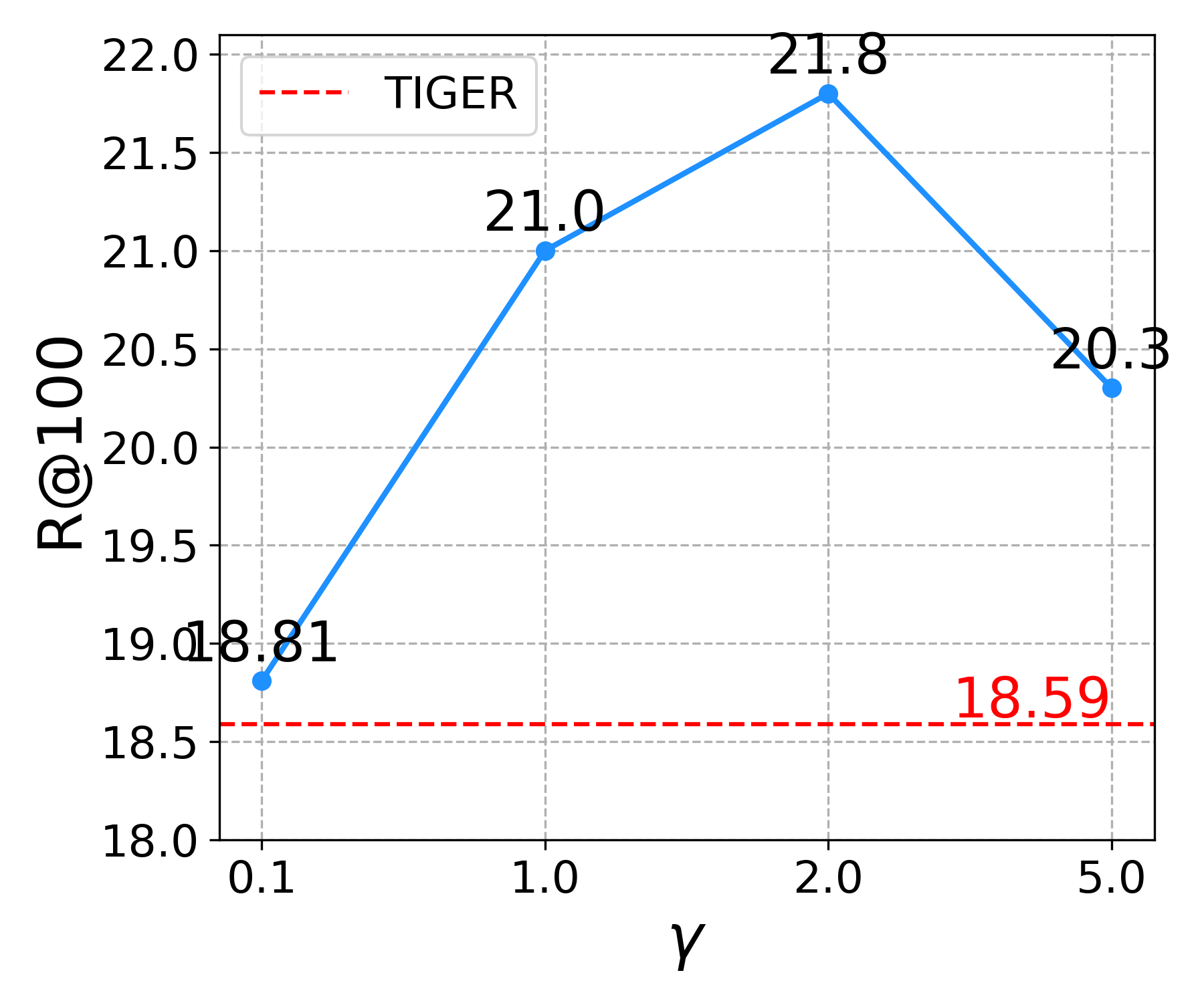}
        \caption{$R@100$ with $\gamma$}
        \label{fig:gamma_100}
        \vspace{-6pt}
    \end{subfigure}
    \hspace{0.04\textwidth}
    \begin{subfigure}[b]{0.20\textwidth}
        \centering
        \includegraphics[width=\textwidth]{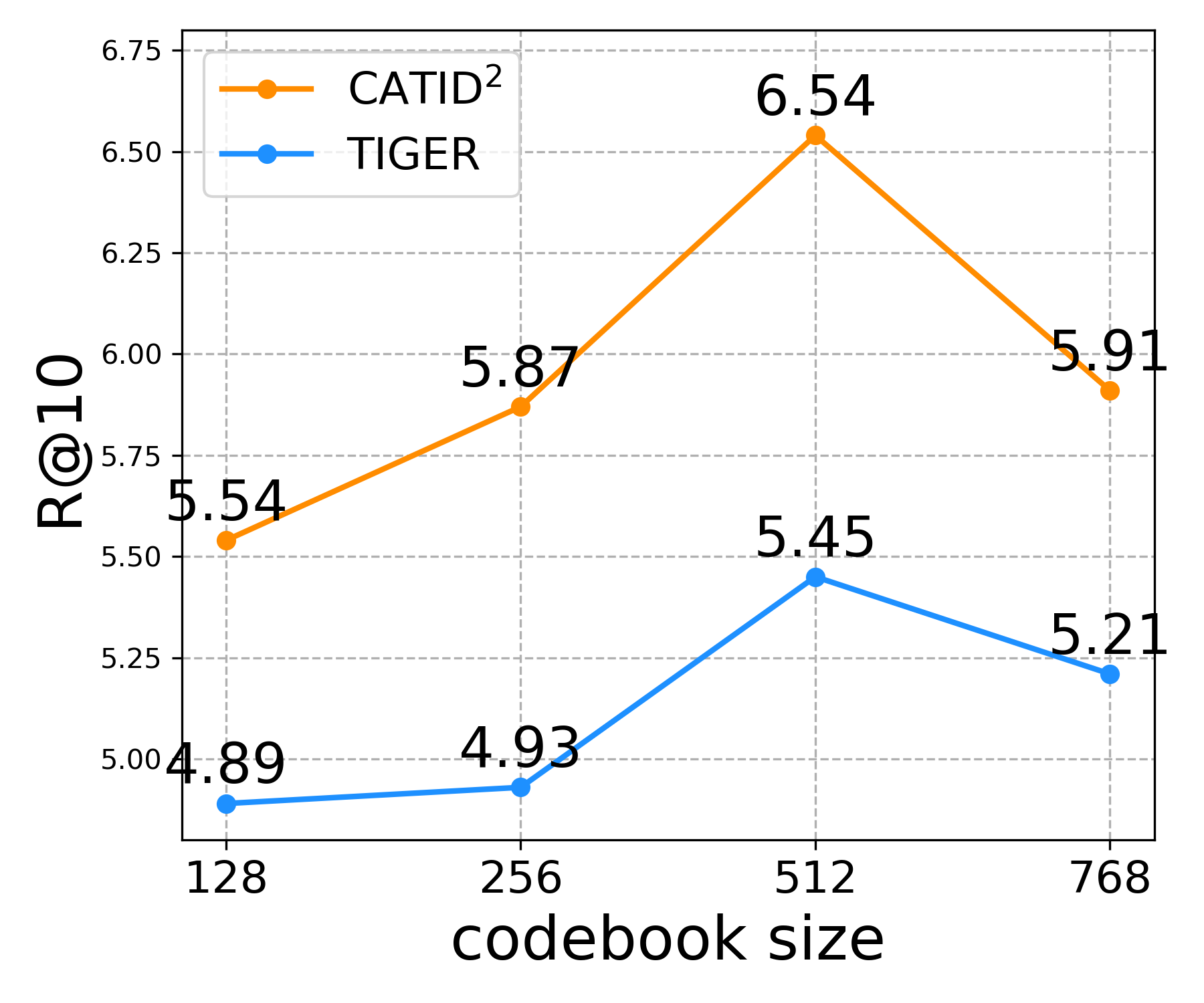}
        \caption{$R@10$ on codebook size}
        \label{fig:codebook}
        \vspace{-6pt}
    \end{subfigure}
    \hspace{0.04\textwidth}
    \begin{subfigure}[b]{0.20\textwidth}
        \centering
        \includegraphics[width=\textwidth]{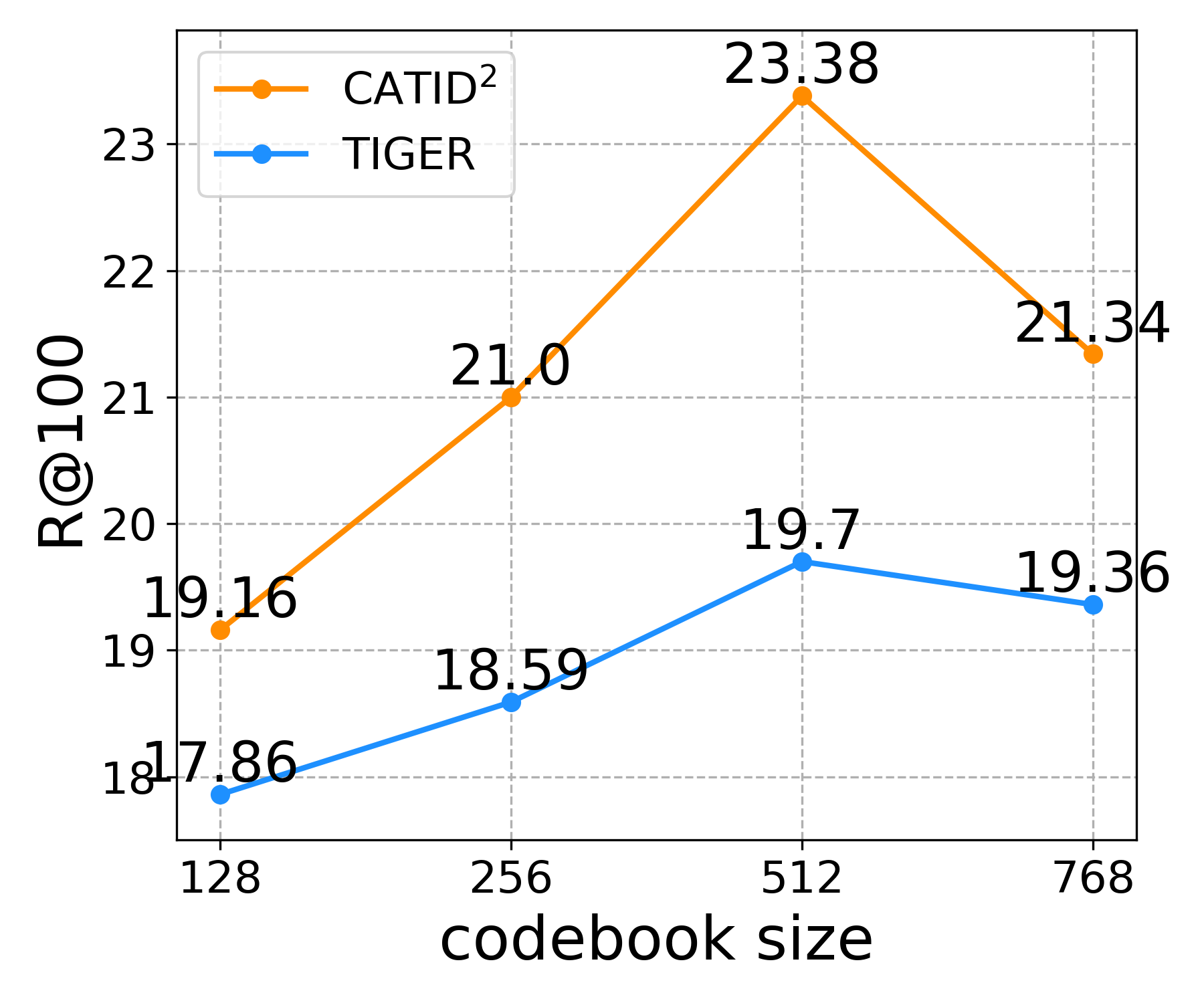}
        \caption{$R@100$ on codebook size}
        \label{fig:codebook}
        \vspace{-6pt}
    \end{subfigure}
    \caption{The impact of different loss weights ($\alpha$, $\beta$, $\gamma$), codebook size, and category depth on model performance. }
    \vspace{-8pt}
    \label{fig:hyper_parameter}
\end{figure*}

\subsection{Hyperparameter Analysis (RQ4)}


\noindent\textbf{Category Depth}
We first analyze the impact of category depth, as shown in Table~\ref{depth}. Here, "1 layer", "2 layer", and "3 layer" refer to using category information with 1, 2, and 3 layers of depth, respectively. The "2\&3 layer" setting refers to randomly selecting half of the third-layer categories, masking their third-layer information, and merging them using their corresponding second-layer category. The results demonstrate that model performance improves as the category depth increases, highlighting the advantage of utilizing a deeper category hierarchy. \textbf{The performance of the "2\&3 layer" setting also indicates that our method remains effective even when the category paths are of unequal length, not just when all categories have the same depth}. 

\noindent\textbf{Sampling Method} 
Furthermore, compare our hard negative sampling strategy against a global negative sampling baseline. In our approach, hard negatives are sourced from the same parent category as the positive sample. This method proves to be markedly superior. \textit{The rationale is that distinguishing between a positive sample and its highly similar hard negatives compels the model to learn a more refined and robust feature space}. Consequently, this process enhances the separability between different categories.

\begin{figure*}[htbp]
    \centering
    \begin{subfigure}[b]{0.205\textwidth}
        \centering
        \includegraphics[width=\textwidth]{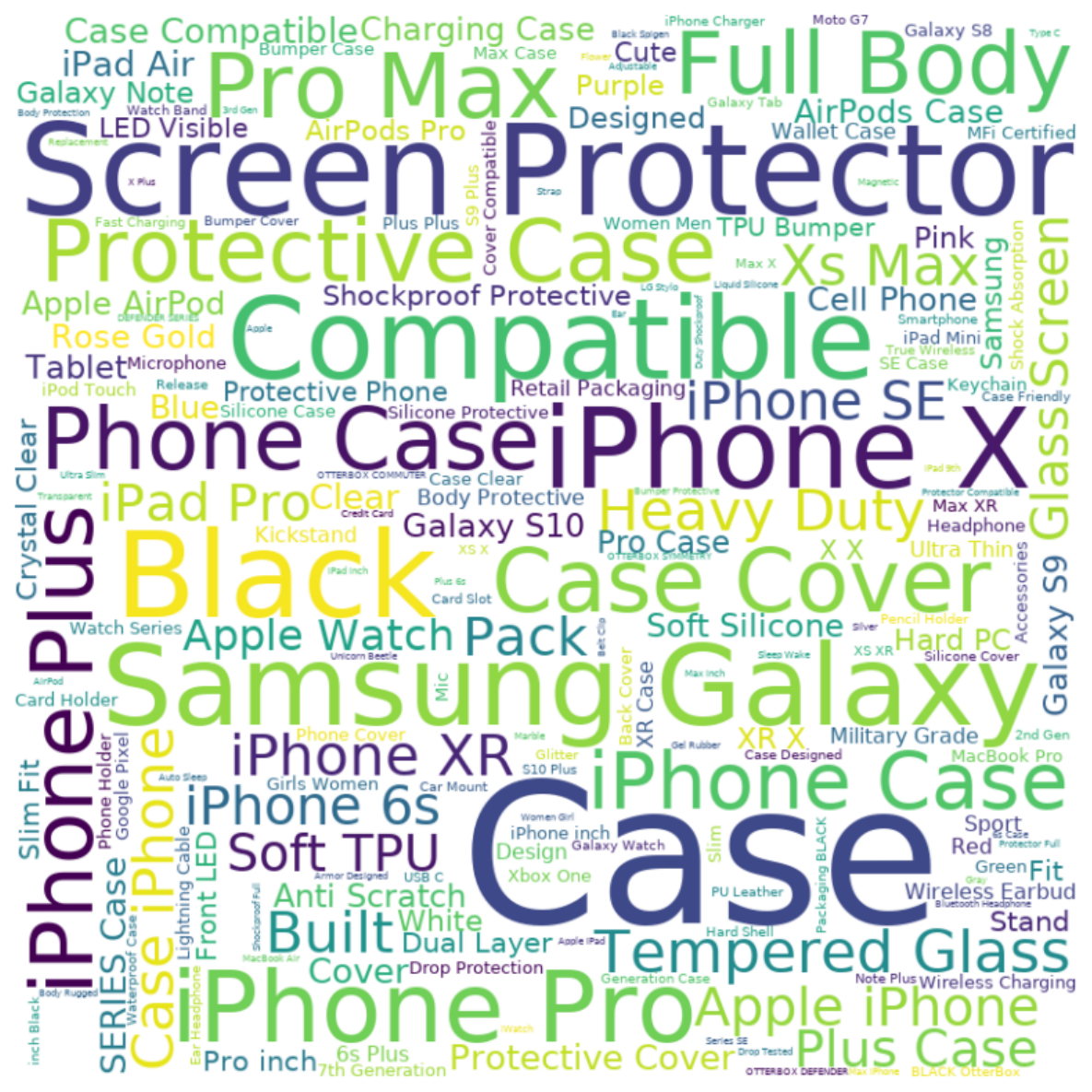}
        \caption{First ID:\texttt{<a\_418>}}
        \label{fig:1}
    \end{subfigure}
    \hspace{0.05\textwidth}
    \begin{subfigure}[b]{0.205\textwidth}
        \centering
        \includegraphics[width=\textwidth]{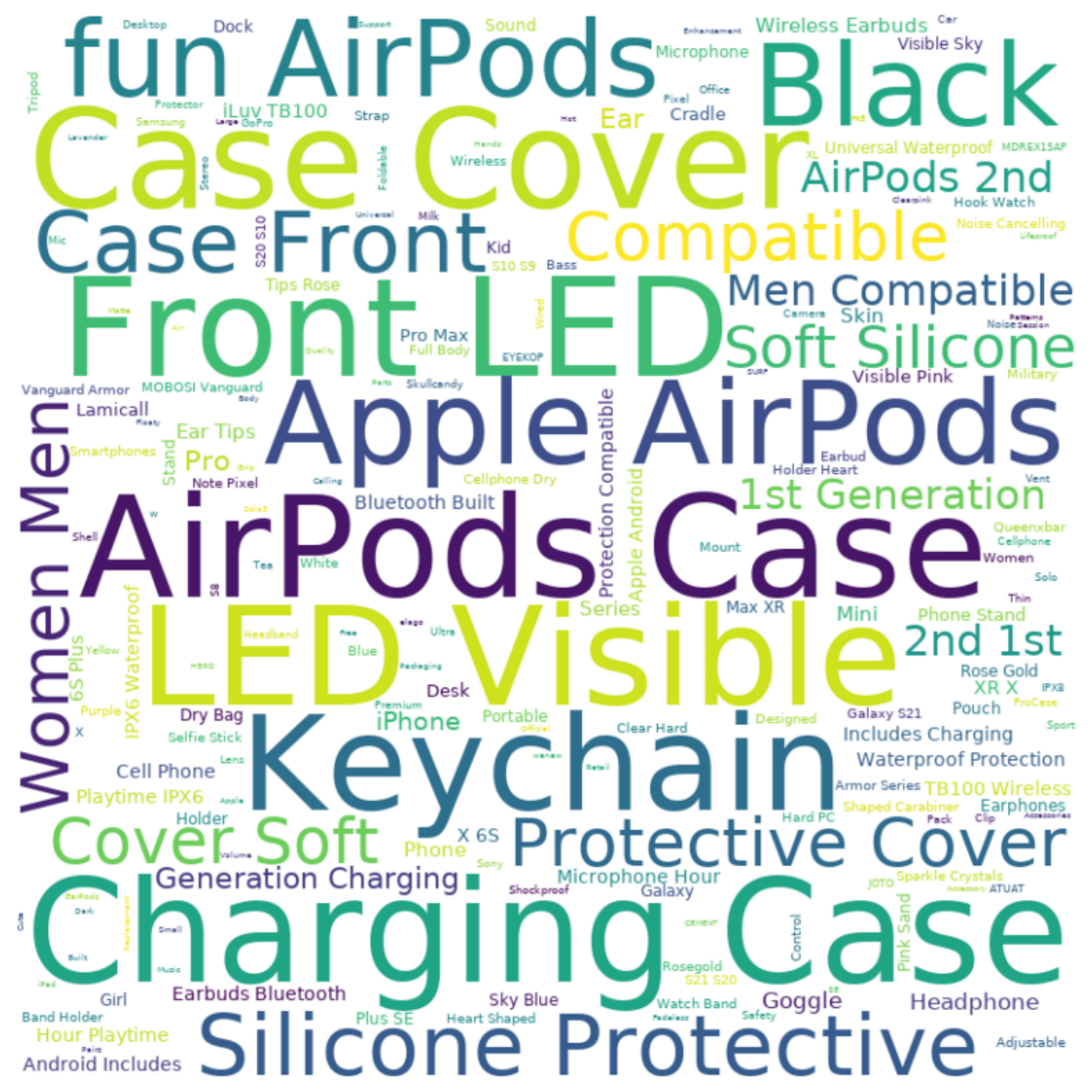}
        \caption{Second ID:\texttt{<a\_418><b\_54>}}
        \label{fig:2}
    \end{subfigure}
    \hspace{0.05\textwidth}
    \begin{subfigure}[b]{0.205\textwidth}
        \centering
        \includegraphics[width=\textwidth]{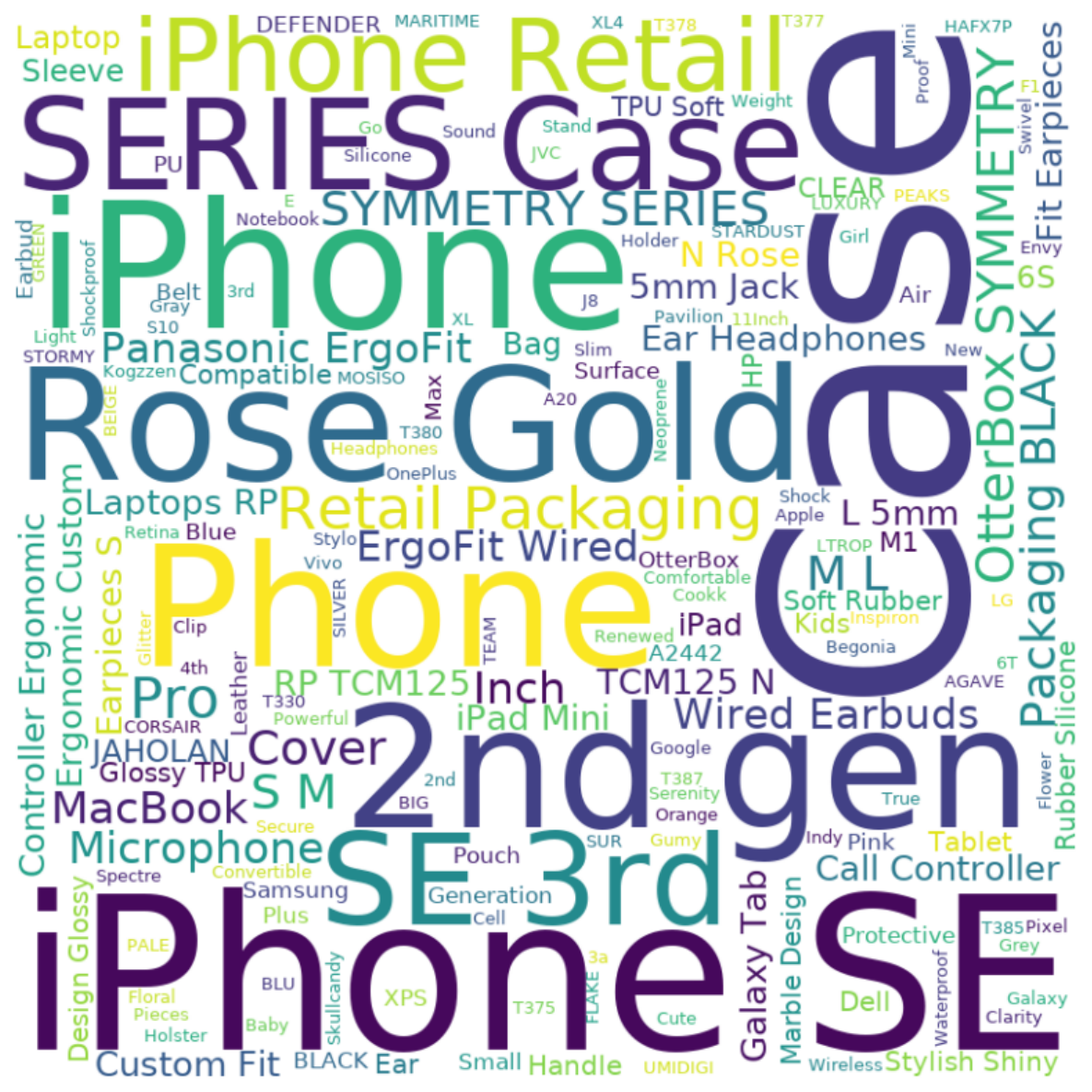}
        \caption{Second ID:\texttt{<a\_418><b\_509>}}
        \label{fig:3}
    \end{subfigure}


    \begin{subfigure}[b]{0.205\textwidth}
        \centering
        \includegraphics[width=\textwidth]{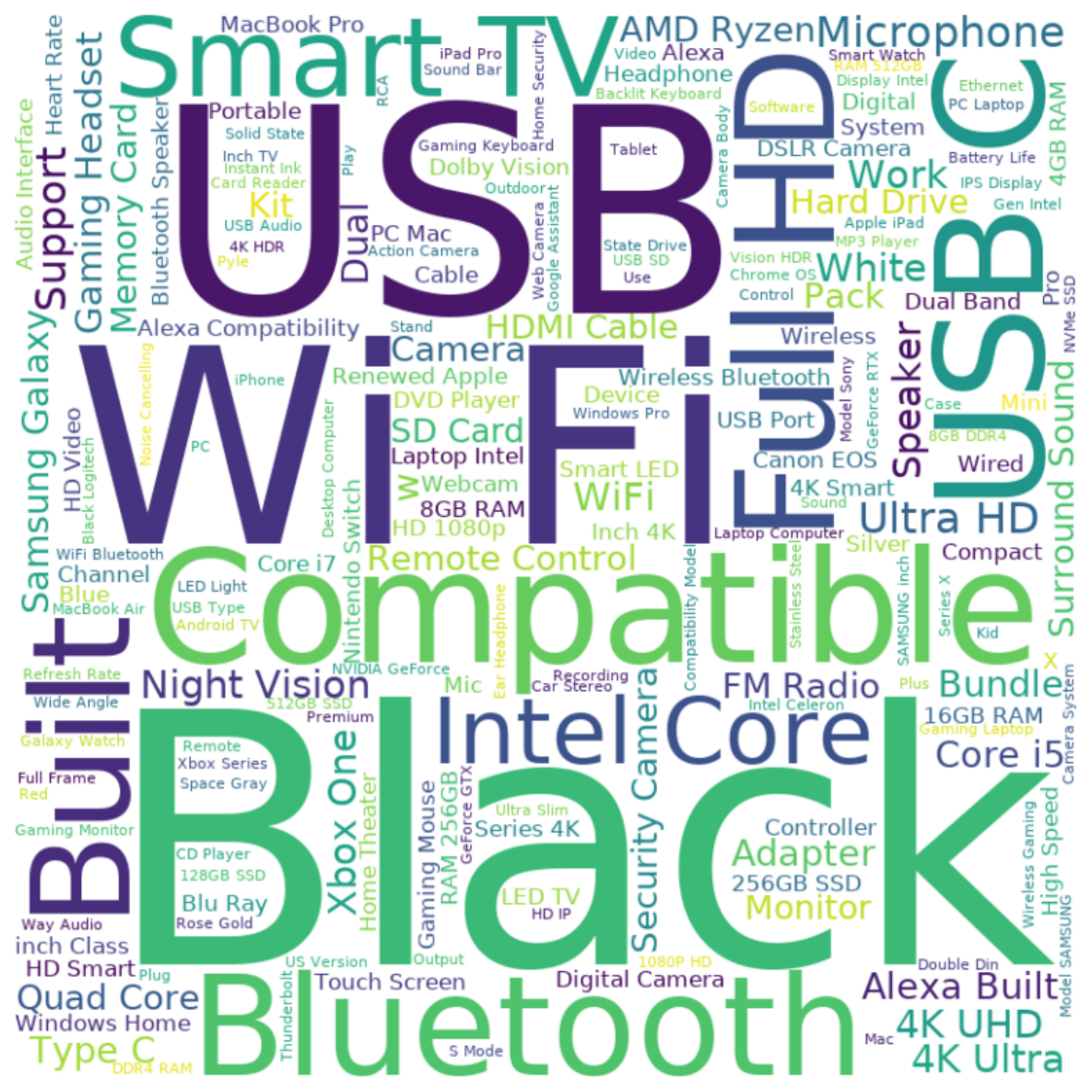}
        \caption{First ID:\texttt{<a\_312>}}
        \label{fig:7}
    \end{subfigure}
    \vspace{-3pt}
    \hspace{0.05\textwidth}
    \begin{subfigure}[b]{0.205\textwidth}
        \centering
        \includegraphics[width=\textwidth]{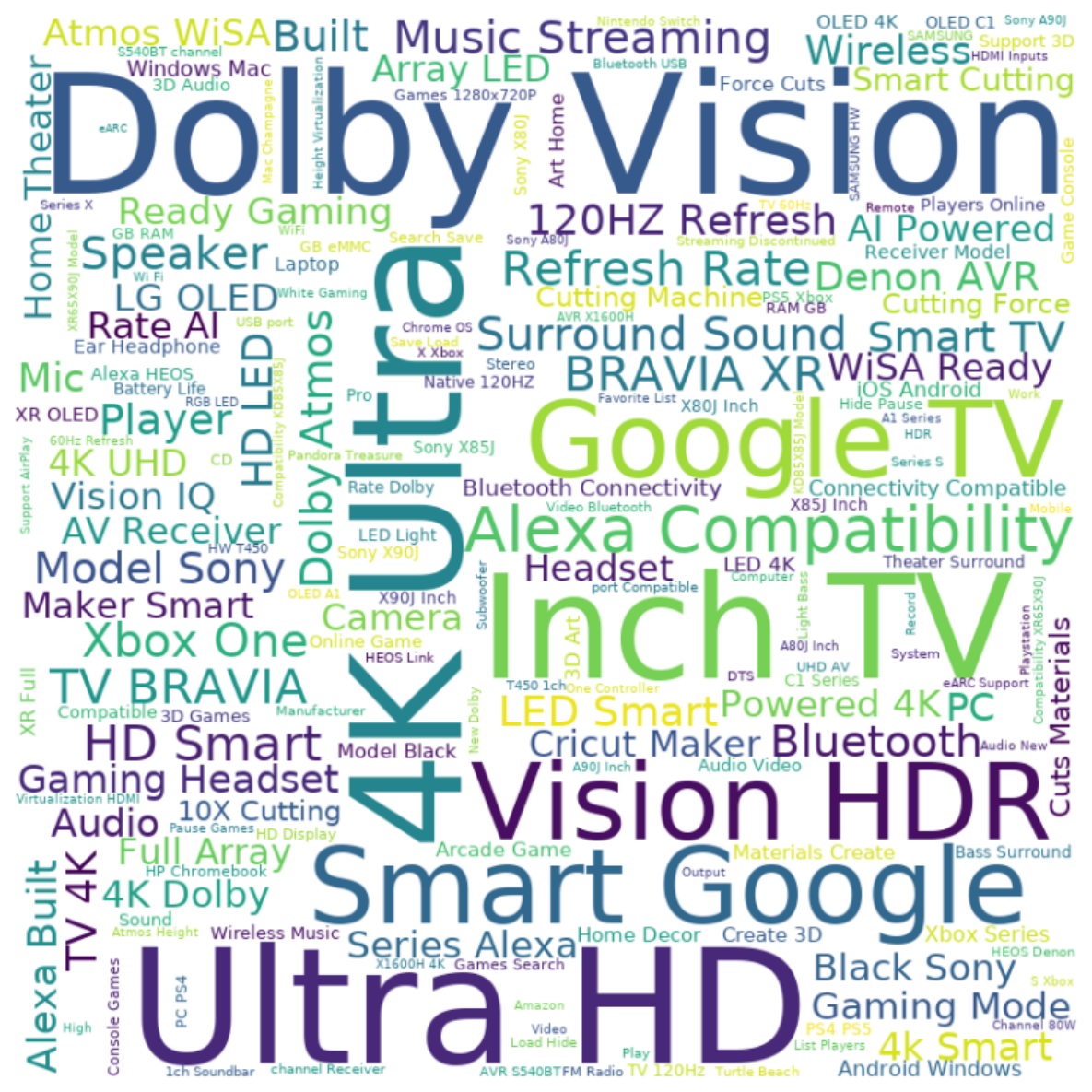}
        \caption{Second ID:\texttt{<a\_312><b\_229>}}
        \label{fig:8}
    \end{subfigure}
    \vspace{-3pt}
    \hspace{0.05\textwidth}
    \begin{subfigure}[b]{0.205\textwidth}
        \centering
        \includegraphics[width=\textwidth]{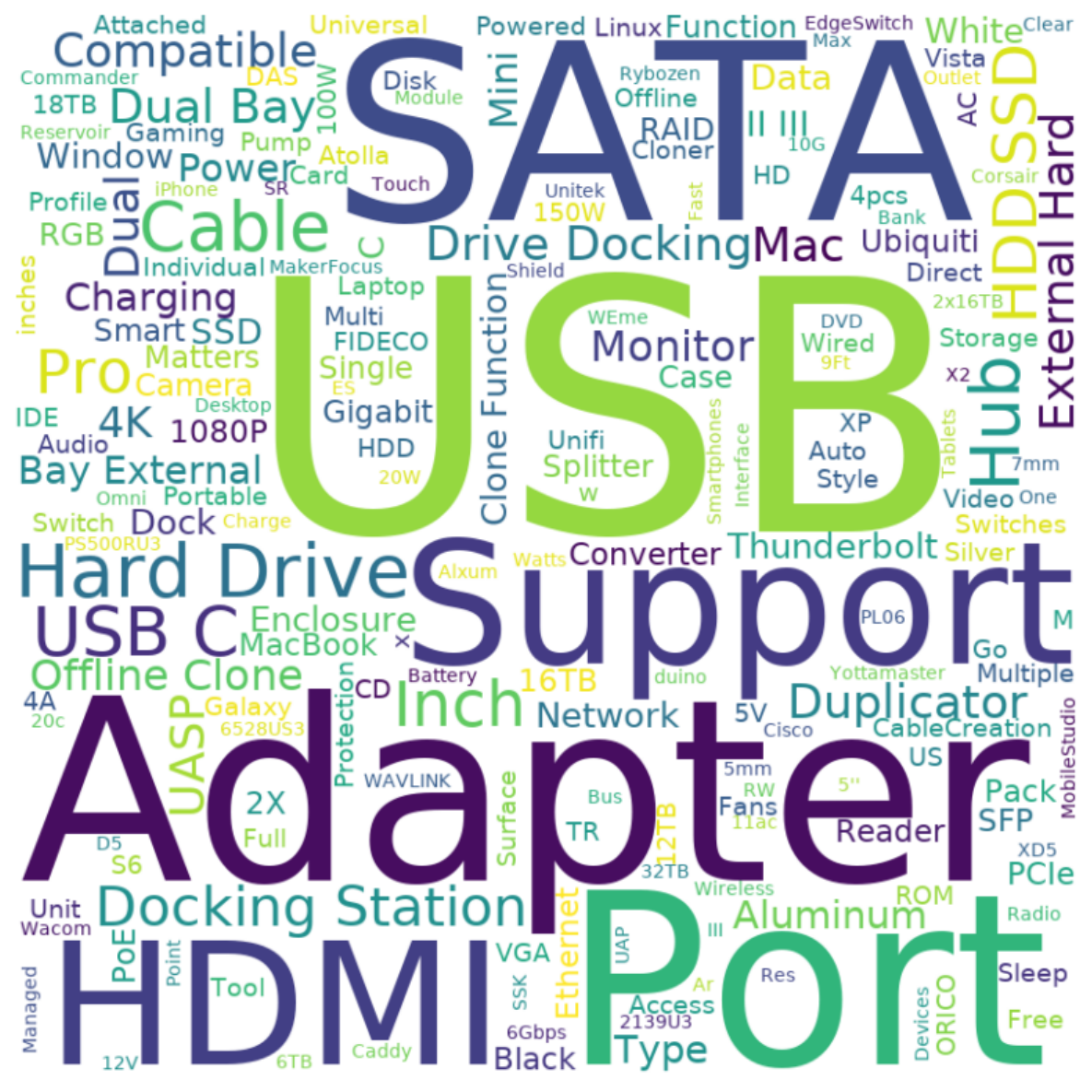}
        \caption{Second ID:\texttt{<a\_312><b\_454>}}
        \label{fig:9}
    \end{subfigure}
    \vspace{-3pt}
    \caption{Case studies of word clouds on the generative DocIDs.}
    \label{fig:wordcloud}
\end{figure*}

\noindent\textbf{Layers of RQ-VAE}
We examine the effect of varying the number of RQ-VAE layers, as presented in Table~\ref{layers}. Both TIGER and CAT-ID$^2$ show improved performance with more layers initially, followed by a decline. We attribute this trend to the following reasons: 
1) With too few RQ-VAE layers, the codebook representation space is limited (e.g., $256^3$), resulting in higher codebook collisions and reduced discriminative power of the generated IDs.
2) Conversely, with too many layers, the representation space becomes excessively sparse (e.g., $256^5$), which hampers retrieval performance due to underutilization of the codebook and increases memory overhead for LLMs because of longer IDs.
Crucially, CAT-ID$^2$ consistently outperforms TIGER across all metrics, regardless of the number of layers, highlighting its robustness and effectiveness.

\noindent\textbf{Loss Weights}
The effects of different loss weights on $R@10$ and $R@100$ are illustrated in Figure~\ref{fig:hyper_parameter}. The weights $\alpha$, $\beta$, and $\gamma$ correspond to the Dispersion Loss, Hierarchical Class Constraint Loss, and Cluster Scale Constraint Loss, respectively. As shown in Figures~\ref{fig:alpha_10} and~\ref{fig:alpha_100}, increasing $\alpha$ initially improves the performance, but after reaching a certain point, the performance begins to decline. This occurs because an excessively large $\alpha$ forces the document representations to cluster too separately, disrupting their inherent relationships. On the other hand, setting $\alpha$ too small reduces the discriminability among documents, leading to performance degradation. Figures~\ref{fig:beta_10} and~\ref{fig:beta_100} show the results under different levels of category constraints, which similarly exhibit an initial increase in performance followed by a decline. 
We hypothesize that when $\beta$ is too large, the constraints are too strict, so the model just follows the original category tree and cannot learn extra information.
Conversely, when $\beta$ is too small, the model introduces insufficient category information, resulting in suboptimal performance. The same reasoning applies to $\gamma$, as shown in Figures~\ref{fig:gamma_10} and~\ref{fig:gamma_100}. When $\gamma$ is too large, the codebook approaches an almost uniform distribution. On the other hand, when $\gamma$ is too small, it fails to balance the risk of collapse introduced by the category constraint.

\noindent\textbf{Codebook Size}
We conduct experiments on the codebook size for both TIGER and CAT-ID$^2$, as shown in Figure~\ref{fig:codebook}. The performance of both models initially improves as the codebook size increases but then begins to decline. This is because a codebook that is too small lacks sufficient discriminability for ID differentiation, while an overly large codebook leads to overly dispersed distributions, which harms model performance. Notably, CAT-ID$^2$ consistently outperforms TIGER across different codebook sizes.

\begin{table}[t]

\renewcommand{\arraystretch}{1.05}
    \centering
    \caption{Comparison of different models over training time. Total time (Hours) includes the duration for both DocID learning and generative model training.}
    \vspace{-4pt}
    \begin{tabular}{c|c}
    \toprule
        Model & Total Training Time \\
    \midrule
       DSI\textsubscript{naive} & 28.2H (one A100 80G)\\
       DSI\textsubscript{semantic} & 15.7H (one A100 80G)\\
       TIGER  & 19.9H (one A100 80G)\\
       NCI  & 32.5H (one A100 80G)\\
       GenRet & 35.1H (\textbf{eight} A100 80G)\\
    \midrule
       CAT-ID$^2$  & 24.8H (one A100 80G)\\
    \bottomrule
       
    \end{tabular}
    \vspace{-8pt}
    \label{tab:time}
\end{table}
\subsection{Training Efficiency (RQ5)}
Training efficiency is critical in industrial applications. To evaluate this, we compare the training time of different models in Table~\ref{tab:time}. As shown, the introduction of contrastive learning does lead to an increase in complexity compared to TIGER. However, the majority of the time consumption is concentrated on the training of the generative model, and the additional time required constitutes only a small fraction of the total duration. This increase is acceptable for practical applications.

\begin{table*}[t]
\small
\centering
\caption{Comparison of generated DocIDs for the items with the same category using CAT-ID$^2$ and TIGER.}

\renewcommand{\arraystretch}{1.13} 
\setlength{\tabcolsep}{8pt} 
\begin{tabular}{|c|c|c|>{\centering\arraybackslash}m{3.4cm}|>{\centering\arraybackslash}m{3.6cm}|>{\centering\arraybackslash}m{3.6cm}|}
\hline
\centering\textbf{Level 1} & \centering\textbf{Level 2} & \centering\textbf{Level 3} & \centering\arraybackslash\textbf{Title} & \centering\arraybackslash\textbf{semantic id (CAT-ID$^2$)} & \centering\arraybackslash\textbf{semantic id (TIGER)} \\
\hline
\multirow{20}{*}{\centering Women} & \multirow{15}{*}{\centering\makecell{Clothing,\\ Shoes \\ \& \\Jewelry}} & \multirow{5}{*}{\centering Shoes} & \centering\arraybackslash adidas Women's Ultraboost Running Shoe & \centering\arraybackslash{\color{red}<a\_65><b\_225>}<c\_205><d\_19> & \centering\arraybackslash{\color{black}<a\_165><b\_33>}<c\_203><d\_133> \\
\cline{4-6}
 &  &  & \centering\arraybackslash Nike Women's Downshifter 9 Sneaker & \centering\arraybackslash{\color{red}<a\_65><b\_225>}<c\_32><d\_0> & \centering\arraybackslash{\color{black}<a\_110><b\_167>}<c\_177><d\_164> \\
\cline{4-6}
 &  &  & \centering\arraybackslash DailyShoes Women's Embroidered Legend Toe... & \centering\arraybackslash{\color{red}<a\_65>}<b\_182><c\_67><d\_19> & \centering\arraybackslash{\color{black}<a\_131><b\_50>}<c\_124><d\_23> \\
\cline{3-6}
 &  & \multirow{3}{*}{\centering Clothing} & \centering\arraybackslash Milumia Womens Vintage Boho Button Up Split Floral... & \centering\arraybackslash{\color{red}<a\_65><b\_182>}<c\_205><d\_30> & \centering\arraybackslash{\color{black}<a\_196><b\_204>}<c\_158><d\_22> \\
\cline{4-6}
 &  &  & \centering\arraybackslash REORIA Women's Sexy Sleeveless Racer Back... & \centering\arraybackslash{\color{red}<a\_65><b\_182>}<c\_116><d\_238> & \centering\arraybackslash{\color{black}<a\_106><b\_221>}<c\_121><d\_20> \\
\cline{3-6}
 &  & \multirow{3}{*}{\centering Accessories} & \centering\arraybackslash Costa Del Mar Women's Panga Square Sunglasses & \centering\arraybackslash{\color{red}<a\_65>}<b\_225><c\_186><d\_105> & \centering\arraybackslash{\color{black}<a\_165><b\_22>}<c\_135><d\_2> \\
\cline{4-6}
 &  &  & \centering\arraybackslash WERFORU No Buckle Stretch Belt For Women... & \centering\arraybackslash{\color{red}<a\_65>}<b\_181><c\_230><d\_4> & \centering\arraybackslash{\color{black}<a\_131><b\_142>}<c\_39><d\_163> \\
\cline{3-6}
 &  & \multirow{3}{*}{\centering\makecell{Handbags\\ \& \\ Wallets}} & \centering\arraybackslash Chala Crossbody Cell Phone Purse - Women... & \centering\arraybackslash{\color{red}<a\_65>}<b\_140><c\_230><d\_0> & \centering\arraybackslash{\color{black}<a\_172><b\_154>}<c\_166><d\_63> \\
\cline{4-6}
 &  &  & \centering\arraybackslash Nylon Travel Tote Cross-body Carry On Bag... & \centering\arraybackslash{\color{red}<a\_65>}<b\_200><c\_46><d\_96> & \centering\arraybackslash{\color{blue}<a\_36>}<b\_85><c\_112><d\_170> \\
\cline{2-6}
 & \multirow{3}{*}{\centering\makecell{Sports\\ \& \\ Outdoors}} & \multirow{3}{*}{\centering Clothing} & \centering\arraybackslash Womens High Waist Yoga Pants Tummy Control... & \centering\arraybackslash{\color{red}<a\_65><b\_120>}<c\_227><d\_9> & \centering\arraybackslash{\color{blue}<a\_36>}<b\_188><c\_127><d\_81> \\
\cline{4-6}
 &  &  & \centering\arraybackslash DILANNI Women's High Waist Yoga Shorts... & \centering\arraybackslash{\color{red}<a\_65><b\_120>}<c\_124><d\_0> & \centering\arraybackslash{\color{blue}<a\_36>}<b\_169><c\_121><d\_0> \\
\hline
\end{tabular}
\label{case}
\end{table*}

\begin{table}[t] 
\renewcommand{\arraystretch}{1.05}
    \centering
    \setlength\tabcolsep{7pt}
   \caption{Results of online A/B Test. We present the relative increase to the online production system.} 
    \begin{tabular}{c|c}
    \toprule
    Query & Ave. Orders\\
    \midrule
    Ambiguous & $+0.33\%$ \\
    Precise\_1 & $+0.08\%$ \\
    Precise\_2 & $+0.10\%$ \\
    Long-tail & $+0.24\%$ \\
    \midrule
    Overall   & $+0.13\%$ \\
    \bottomrule
    \end{tabular}
    \vspace{-6pt}
    \label{tab:A/B}
\end{table}

\subsection{Online A/B Test (RQ6)}
CAT-ID\(^2\) was deployed in the recall stage of our online search system for a 10-day A/B test on the e-commerce platform. To address latency constraints, we combined methods from Hi-Gen~\cite{hi-gen} and GDR~\cite{gdr}, truncating results based on intermediate node clusters and using Dense Retrieval (DR) scores as the truncation criterion. The results are summarized in Table~\ref{tab:A/B}.

The first column shows different query intents, with actual intent names anonymized for confidentiality. Query intents were determined using a rule-based approach after named entity recognition, following the priority order: Ambiguous → Precise\_1 → Precise\_2 → Long-tail. The key metric is the average number of orders per thousand users, which is critical in our business, where even a 0.1\% improvement is significant. As shown in the table, compared to our existing recall system, CAT-ID$^2$ performed particularly well on Ambiguous and Long-tail queries, with improvements of 0.33\% and 0.24\%, respectively. For Precise\_1 and Precise\_2 queries, the gains were smaller but still meaningful, at +0.08\% and +0.10\%. Overall, CAT-ID$^2$ increased average orders by +0.13\%.

It is worth noting that our recall system integrates multiple retrieval methods, including sparse and dense retrieval and LLM-based query rewriting. On this foundation, our GR model delivered better performance, demonstrating that the end-to-end GR paradigm effectively mitigates information loss caused by traditional decoupled query understanding. This highlights the strong potential of end-to-end GR as an emerging approach in real-world retrieval scenarios.

\subsection{Case Study}
\noindent\textbf{Word clouds of product titles.}
Figure~\ref{fig:wordcloud} uses word clouds to illustrate the hierarchical nature of our DocIDs. For instance, the top-level ID <a\_31>(Figure~\ref{fig:7}) broadly covers connectivity devices ("USB," "Wi-Fi"). Its sub-IDs refine this category into smart TVs with "Dolby Vision" (<a\_31><b\_229>, Figure~\ref{fig:8}) and hardware adapters supporting "SATA" and "HDMI" (<a\_31><b\_454>, Figure~\ref{fig:9}). Similarly, ID <a\_418> (Figure~\ref{fig:1}) represents protective cases ("iPhone," "Case"). This is further specialized into sub-IDs for AirPods cases ( <a\_418><b\_54>, Figure~\ref{fig:2}) and iPhone cases (<a\_418><b\_509>, Figure~\ref{fig:3}). These examples show that DocIDs effectively map the product hierarchy.

\noindent\textbf{DocIDs Comparison.}
We compared DocIDs from CAT-ID$^2$ against those from TIGER in Table~\ref{case}. The common prefixes for each model are highlighted in {\color{red}red} (CAT-ID$^2$) and {\color{blue}blue} (TIGER). It can be observed that the DocIDs generated by CAT-ID$^2$ for documents within the same category have a longer common prefix. In contrast, TIGER performs poorly, often failing to assign even the same top-level ID to similar products. This indicates CAT-ID$^2$'s enhanced capability in capturing and representing category structure.
\section{Conclusion}
In this paper, we proposed CAT-ID$^2$, a method that effectively utilizes Category-Tree information for hierarchical semantic ID construction while preserving global semantic context. CAT-ID$^2$ integrates three key loss functions: the Hierarchical Class Constraint Loss, which reflects the hierarchical structure by aligning representations within categories and separating those across categories; the Cluster Scale Constraint Loss, which prevents encoding collapse by balancing category sizes; and the Dispersion Loss, which ensures uniqueness and diversity of the IDs. By combining these components, CAT-ID$^2$ generates semantically meaningful IDs. Both online and offline experiments confirm the effectiveness of CAT-ID$^2$.

\section*{Acknowledgments}
This work was supported by the National Key Research and Development Program of China under Grant No. 2024YFF0729003, the National Natural Science Foundation of China under Grant Nos. 62176014, 62206266, the Fundamental Research Funds for the Central Universities.


\bibliographystyle{ACM-Reference-Format}
\bibliography{wsdm_ref}


\begin{thebibliography}{36}


\ifx \showCODEN    \undefined \def \showCODEN     #1{\unskip}     \fi
\ifx \showISBNx    \undefined \def \showISBNx     #1{\unskip}     \fi
\ifx \showISBNxiii \undefined \def \showISBNxiii  #1{\unskip}     \fi
\ifx \showISSN     \undefined \def \showISSN      #1{\unskip}     \fi
\ifx \showLCCN     \undefined \def \showLCCN      #1{\unskip}     \fi
\ifx \shownote     \undefined \def \shownote      #1{#1}          \fi
\ifx \showarticletitle \undefined \def \showarticletitle #1{#1}   \fi
\ifx \showURL      \undefined \def \showURL       {\relax}        \fi
\providecommand\bibfield[2]{#2}
\providecommand\bibinfo[2]{#2}
\providecommand\natexlab[1]{#1}
\providecommand\showeprint[2][]{arXiv:#2}

\bibitem[Bevilacqua et~al\mbox{.}(2022)]%
        {seal}
\bibfield{author}{\bibinfo{person}{Michele Bevilacqua}, \bibinfo{person}{Giuseppe Ottaviano}, \bibinfo{person}{Patrick Lewis}, \bibinfo{person}{Scott Yih}, \bibinfo{person}{Sebastian Riedel}, {and} \bibinfo{person}{Fabio Petroni}.} \bibinfo{year}{2022}\natexlab{}.
\newblock \showarticletitle{Autoregressive search engines: Generating substrings as document identifiers}.
\newblock \bibinfo{journal}{\emph{Advances in Neural Information Processing Systems}}  \bibinfo{volume}{35} (\bibinfo{year}{2022}), \bibinfo{pages}{31668--31683}.
\newblock


\bibitem[Cuturi(2013)]%
        {cuturi2013sinkhorn}
\bibfield{author}{\bibinfo{person}{Marco Cuturi}.} \bibinfo{year}{2013}\natexlab{}.
\newblock \showarticletitle{Sinkhorn distances: Lightspeed computation of optimal transport}.
\newblock \bibinfo{journal}{\emph{Advances in neural information processing systems}}  \bibinfo{volume}{26} (\bibinfo{year}{2013}).
\newblock


\bibitem[Devlin et~al\mbox{.}(2019)]%
        {bert}
\bibfield{author}{\bibinfo{person}{Jacob Devlin}, \bibinfo{person}{Ming-Wei Chang}, \bibinfo{person}{Kenton Lee}, {and} \bibinfo{person}{Kristina Toutanova}.} \bibinfo{year}{2019}\natexlab{}.
\newblock \showarticletitle{Bert: Pre-training of deep bidirectional transformers for language understanding}. In \bibinfo{booktitle}{\emph{Proceedings of the 2019 conference of the North American chapter of the association for computational linguistics: human language technologies, volume 1 (long and short papers)}}. \bibinfo{pages}{4171--4186}.
\newblock


\bibitem[Hofst{\"a}tter et~al\mbox{.}(2021)]%
        {kd}
\bibfield{author}{\bibinfo{person}{Sebastian Hofst{\"a}tter}, \bibinfo{person}{Sheng-Chieh Lin}, \bibinfo{person}{Jheng-Hong Yang}, \bibinfo{person}{Jimmy Lin}, {and} \bibinfo{person}{Allan Hanbury}.} \bibinfo{year}{2021}\natexlab{}.
\newblock \showarticletitle{Efficiently teaching an effective dense retriever with balanced topic aware sampling}. In \bibinfo{booktitle}{\emph{Proceedings of the 44th International ACM SIGIR Conference on Research and Development in Information Retrieval}}. \bibinfo{pages}{113--122}.
\newblock


\bibitem[Karpukhin et~al\mbox{.}(2020)]%
        {dpr}
\bibfield{author}{\bibinfo{person}{Vladimir Karpukhin}, \bibinfo{person}{Barlas Oguz}, \bibinfo{person}{Sewon Min}, \bibinfo{person}{Patrick Lewis}, \bibinfo{person}{Ledell Wu}, \bibinfo{person}{Sergey Edunov}, \bibinfo{person}{Danqi Chen}, {and} \bibinfo{person}{Wen-tau Yih}.} \bibinfo{year}{2020}\natexlab{}.
\newblock \showarticletitle{Dense Passage Retrieval for Open-Domain Question Answering}. In \bibinfo{booktitle}{\emph{Proceedings of the 2020 Conference on Empirical Methods in Natural Language Processing (EMNLP)}}. \bibinfo{pages}{6769--6781}.
\newblock


\bibitem[Lee et~al\mbox{.}(2022)]%
        {rq-image}
\bibfield{author}{\bibinfo{person}{Doyup Lee}, \bibinfo{person}{Chiheon Kim}, \bibinfo{person}{Saehoon Kim}, \bibinfo{person}{Minsu Cho}, {and} \bibinfo{person}{Wook-Shin Han}.} \bibinfo{year}{2022}\natexlab{}.
\newblock \showarticletitle{Autoregressive image generation using residual quantization}. In \bibinfo{booktitle}{\emph{Proceedings of the IEEE/CVF Conference on Computer Vision and Pattern Recognition}}. \bibinfo{pages}{11523--11532}.
\newblock


\bibitem[Liu and Mozafari(2024)]%
        {liu2024query}
\bibfield{author}{\bibinfo{person}{Jie Liu} {and} \bibinfo{person}{Barzan Mozafari}.} \bibinfo{year}{2024}\natexlab{}.
\newblock \showarticletitle{Query rewriting via large language models}.
\newblock \bibinfo{journal}{\emph{arXiv preprint arXiv:2403.09060}} (\bibinfo{year}{2024}).
\newblock


\bibitem[Nguyen and Yates(2023)]%
        {dense_1}
\bibfield{author}{\bibinfo{person}{Thong Nguyen} {and} \bibinfo{person}{Andrew Yates}.} \bibinfo{year}{2023}\natexlab{}.
\newblock \showarticletitle{Generative retrieval as dense retrieval}.
\newblock \bibinfo{journal}{\emph{arXiv preprint arXiv:2306.11397}} (\bibinfo{year}{2023}).
\newblock


\bibitem[Ni et~al\mbox{.}(2022)]%
        {sentence-t5}
\bibfield{author}{\bibinfo{person}{Jianmo Ni}, \bibinfo{person}{Gustavo~Hernandez Abrego}, \bibinfo{person}{Noah Constant}, \bibinfo{person}{Ji Ma}, \bibinfo{person}{Keith Hall}, \bibinfo{person}{Daniel Cer}, {and} \bibinfo{person}{Yinfei Yang}.} \bibinfo{year}{2022}\natexlab{}.
\newblock \showarticletitle{Sentence-T5: Scalable Sentence Encoders from Pre-trained Text-to-Text Models}. In \bibinfo{booktitle}{\emph{Findings of the Association for Computational Linguistics: ACL 2022}}. \bibinfo{pages}{1864--1874}.
\newblock


\bibitem[Oord et~al\mbox{.}(2018)]%
        {infonce}
\bibfield{author}{\bibinfo{person}{Aaron van~den Oord}, \bibinfo{person}{Yazhe Li}, {and} \bibinfo{person}{Oriol Vinyals}.} \bibinfo{year}{2018}\natexlab{}.
\newblock \showarticletitle{Representation learning with contrastive predictive coding}.
\newblock \bibinfo{journal}{\emph{arXiv preprint arXiv:1807.03748}} (\bibinfo{year}{2018}).
\newblock


\bibitem[Peng et~al\mbox{.}(2024)]%
        {query_rewriting}
\bibfield{author}{\bibinfo{person}{Wenjun Peng}, \bibinfo{person}{Guiyang Li}, \bibinfo{person}{Yue Jiang}, \bibinfo{person}{Zilong Wang}, \bibinfo{person}{Dan Ou}, \bibinfo{person}{Xiaoyi Zeng}, \bibinfo{person}{Derong Xu}, \bibinfo{person}{Tong Xu}, {and} \bibinfo{person}{Enhong Chen}.} \bibinfo{year}{2024}\natexlab{}.
\newblock \showarticletitle{Large language model based long-tail query rewriting in taobao search}. In \bibinfo{booktitle}{\emph{Companion Proceedings of the ACM Web Conference 2024}}. \bibinfo{pages}{20--28}.
\newblock


\bibitem[Qu et~al\mbox{.}(2021)]%
        {hard_mining}
\bibfield{author}{\bibinfo{person}{Yingqi Qu}, \bibinfo{person}{Yuchen Ding}, \bibinfo{person}{Jing Liu}, \bibinfo{person}{Kai Liu}, \bibinfo{person}{Ruiyang Ren}, \bibinfo{person}{Wayne~Xin Zhao}, \bibinfo{person}{Daxiang Dong}, \bibinfo{person}{Hua Wu}, {and} \bibinfo{person}{Haifeng Wang}.} \bibinfo{year}{2021}\natexlab{}.
\newblock \showarticletitle{RocketQA: An Optimized Training Approach to Dense Passage Retrieval for Open-Domain Question Answering}. In \bibinfo{booktitle}{\emph{Proceedings of the 2021 Conference of the North American Chapter of the Association for Computational Linguistics: Human Language Technologies}}. \bibinfo{pages}{5835--5847}.
\newblock


\bibitem[Raffel et~al\mbox{.}(2020)]%
        {t5}
\bibfield{author}{\bibinfo{person}{Colin Raffel}, \bibinfo{person}{Noam Shazeer}, \bibinfo{person}{Adam Roberts}, \bibinfo{person}{Katherine Lee}, \bibinfo{person}{Sharan Narang}, \bibinfo{person}{Michael Matena}, \bibinfo{person}{Yanqi Zhou}, \bibinfo{person}{Wei Li}, {and} \bibinfo{person}{Peter~J Liu}.} \bibinfo{year}{2020}\natexlab{}.
\newblock \showarticletitle{Exploring the limits of transfer learning with a unified text-to-text transformer}.
\newblock \bibinfo{journal}{\emph{Journal of machine learning research}} \bibinfo{volume}{21}, \bibinfo{number}{140} (\bibinfo{year}{2020}), \bibinfo{pages}{1--67}.
\newblock


\bibitem[Rajput et~al\mbox{.}(2023a)]%
        {rqvae}
\bibfield{author}{\bibinfo{person}{Shashank Rajput}, \bibinfo{person}{Nikhil Mehta}, \bibinfo{person}{Anima Singh}, \bibinfo{person}{Raghunandan Hulikal~Keshavan}, \bibinfo{person}{Trung Vu}, \bibinfo{person}{Lukasz Heldt}, \bibinfo{person}{Lichan Hong}, \bibinfo{person}{Yi Tay}, \bibinfo{person}{Vinh Tran}, \bibinfo{person}{Jonah Samost}, {et~al\mbox{.}}} \bibinfo{year}{2023}\natexlab{a}.
\newblock \showarticletitle{Recommender systems with generative retrieval}.
\newblock \bibinfo{journal}{\emph{Advances in Neural Information Processing Systems}}  \bibinfo{volume}{36} (\bibinfo{year}{2023}), \bibinfo{pages}{10299--10315}.
\newblock


\bibitem[Rajput et~al\mbox{.}(2023b)]%
        {tiger}
\bibfield{author}{\bibinfo{person}{Shashank Rajput}, \bibinfo{person}{Nikhil Mehta}, \bibinfo{person}{Anima Singh}, \bibinfo{person}{Raghunandan Hulikal~Keshavan}, \bibinfo{person}{Trung Vu}, \bibinfo{person}{Lukasz Heldt}, \bibinfo{person}{Lichan Hong}, \bibinfo{person}{Yi Tay}, \bibinfo{person}{Vinh Tran}, \bibinfo{person}{Jonah Samost}, {et~al\mbox{.}}} \bibinfo{year}{2023}\natexlab{b}.
\newblock \showarticletitle{Recommender systems with generative retrieval}.
\newblock \bibinfo{journal}{\emph{Advances in Neural Information Processing Systems}}  \bibinfo{volume}{36} (\bibinfo{year}{2023}), \bibinfo{pages}{10299--10315}.
\newblock


\bibitem[Reddy et~al\mbox{.}(2022)]%
        {esci}
\bibfield{author}{\bibinfo{person}{Chandan~K Reddy}, \bibinfo{person}{Llu{\'\i}s M{\`a}rquez}, \bibinfo{person}{Fran Valero}, \bibinfo{person}{Nikhil Rao}, \bibinfo{person}{Hugo Zaragoza}, \bibinfo{person}{Sambaran Bandyopadhyay}, \bibinfo{person}{Arnab Biswas}, \bibinfo{person}{Anlu Xing}, {and} \bibinfo{person}{Karthik Subbian}.} \bibinfo{year}{2022}\natexlab{}.
\newblock \showarticletitle{Shopping queries dataset: A large-scale ESCI benchmark for improving product search}.
\newblock \bibinfo{journal}{\emph{arXiv preprint arXiv:2206.06588}} (\bibinfo{year}{2022}).
\newblock


\bibitem[Reimers(2019)]%
        {reimers2019sentence}
\bibfield{author}{\bibinfo{person}{N Reimers}.} \bibinfo{year}{2019}\natexlab{}.
\newblock \showarticletitle{Sentence-BERT: Sentence Embeddings using Siamese BERT-Networks}.
\newblock \bibinfo{journal}{\emph{arXiv preprint arXiv:1908.10084}} (\bibinfo{year}{2019}).
\newblock


\bibitem[Robertson et~al\mbox{.}(2009)]%
        {bm25}
\bibfield{author}{\bibinfo{person}{Stephen Robertson}, \bibinfo{person}{Hugo Zaragoza}, {et~al\mbox{.}}} \bibinfo{year}{2009}\natexlab{}.
\newblock \showarticletitle{The probabilistic relevance framework: BM25 and beyond}.
\newblock \bibinfo{journal}{\emph{Foundations and Trends{\textregistered} in Information Retrieval}} \bibinfo{volume}{3}, \bibinfo{number}{4} (\bibinfo{year}{2009}), \bibinfo{pages}{333--389}.
\newblock


\bibitem[Robertson and Walker(1997)]%
        {tf-idf}
\bibfield{author}{\bibinfo{person}{Stephen~E Robertson} {and} \bibinfo{person}{Steve Walker}.} \bibinfo{year}{1997}\natexlab{}.
\newblock \showarticletitle{On relevance weights with little relevance information}. In \bibinfo{booktitle}{\emph{Proceedings of the 20th annual international ACM SIGIR conference on Research and development in information retrieval}}. \bibinfo{pages}{16--24}.
\newblock


\bibitem[Song et~al\mbox{.}(2020)]%
        {song2020mpnet}
\bibfield{author}{\bibinfo{person}{Kaitao Song}, \bibinfo{person}{Xu Tan}, \bibinfo{person}{Tao Qin}, \bibinfo{person}{Jianfeng Lu}, {and} \bibinfo{person}{Tie-Yan Liu}.} \bibinfo{year}{2020}\natexlab{}.
\newblock \showarticletitle{Mpnet: Masked and permuted pre-training for language understanding}.
\newblock \bibinfo{journal}{\emph{Advances in neural information processing systems}}  \bibinfo{volume}{33} (\bibinfo{year}{2020}), \bibinfo{pages}{16857--16867}.
\newblock


\bibitem[Sun et~al\mbox{.}(2024)]%
        {genret}
\bibfield{author}{\bibinfo{person}{Weiwei Sun}, \bibinfo{person}{Lingyong Yan}, \bibinfo{person}{Zheng Chen}, \bibinfo{person}{Shuaiqiang Wang}, \bibinfo{person}{Haichao Zhu}, \bibinfo{person}{Pengjie Ren}, \bibinfo{person}{Zhumin Chen}, \bibinfo{person}{Dawei Yin}, \bibinfo{person}{Maarten Rijke}, {and} \bibinfo{person}{Zhaochun Ren}.} \bibinfo{year}{2024}\natexlab{}.
\newblock \showarticletitle{Learning to tokenize for generative retrieval}.
\newblock \bibinfo{journal}{\emph{Advances in Neural Information Processing Systems}}  \bibinfo{volume}{36} (\bibinfo{year}{2024}).
\newblock


\bibitem[Tang et~al\mbox{.}(2024)]%
        {grgr}
\bibfield{author}{\bibinfo{person}{Yubao Tang}, \bibinfo{person}{Ruqing Zhang}, \bibinfo{person}{Jiafeng Guo}, \bibinfo{person}{Maarten de Rijke}, \bibinfo{person}{Wei Chen}, {and} \bibinfo{person}{Xueqi Cheng}.} \bibinfo{year}{2024}\natexlab{}.
\newblock \showarticletitle{Generative Retrieval Meets Multi-Graded Relevance}. In \bibinfo{booktitle}{\emph{The Thirty-eighth Annual Conference on Neural Information Processing Systems}}.
\newblock


\bibitem[Tay et~al\mbox{.}(2022)]%
        {dsi}
\bibfield{author}{\bibinfo{person}{Yi Tay}, \bibinfo{person}{Vinh Tran}, \bibinfo{person}{Mostafa Dehghani}, \bibinfo{person}{Jianmo Ni}, \bibinfo{person}{Dara Bahri}, \bibinfo{person}{Harsh Mehta}, \bibinfo{person}{Zhen Qin}, \bibinfo{person}{Kai Hui}, \bibinfo{person}{Zhe Zhao}, \bibinfo{person}{Jai Gupta}, {et~al\mbox{.}}} \bibinfo{year}{2022}\natexlab{}.
\newblock \showarticletitle{Transformer memory as a differentiable search index}.
\newblock \bibinfo{journal}{\emph{Advances in Neural Information Processing Systems}}  \bibinfo{volume}{35} (\bibinfo{year}{2022}), \bibinfo{pages}{21831--21843}.
\newblock


\bibitem[Van~der Maaten and Hinton(2008)]%
        {van2008visualizing}
\bibfield{author}{\bibinfo{person}{Laurens Van~der Maaten} {and} \bibinfo{person}{Geoffrey Hinton}.} \bibinfo{year}{2008}\natexlab{}.
\newblock \showarticletitle{Visualizing data using t-SNE.}
\newblock \bibinfo{journal}{\emph{Journal of machine learning research}} \bibinfo{volume}{9}, \bibinfo{number}{11} (\bibinfo{year}{2008}).
\newblock


\bibitem[Wang et~al\mbox{.}(2024)]%
        {letter}
\bibfield{author}{\bibinfo{person}{Wenjie Wang}, \bibinfo{person}{Honghui Bao}, \bibinfo{person}{Xinyu Lin}, \bibinfo{person}{Jizhi Zhang}, \bibinfo{person}{Yongqi Li}, \bibinfo{person}{Fuli Feng}, \bibinfo{person}{See-Kiong Ng}, {and} \bibinfo{person}{Tat-Seng Chua}.} \bibinfo{year}{2024}\natexlab{}.
\newblock \showarticletitle{Learnable item tokenization for generative recommendation}. In \bibinfo{booktitle}{\emph{Proceedings of the 33rd ACM International Conference on Information and Knowledge Management}}. \bibinfo{pages}{2400--2409}.
\newblock


\bibitem[Wang et~al\mbox{.}(2022)]%
        {nci}
\bibfield{author}{\bibinfo{person}{Yujing Wang}, \bibinfo{person}{Yingyan Hou}, \bibinfo{person}{Haonan Wang}, \bibinfo{person}{Ziming Miao}, \bibinfo{person}{Shibin Wu}, \bibinfo{person}{Qi Chen}, \bibinfo{person}{Yuqing Xia}, \bibinfo{person}{Chengmin Chi}, \bibinfo{person}{Guoshuai Zhao}, \bibinfo{person}{Zheng Liu}, {et~al\mbox{.}}} \bibinfo{year}{2022}\natexlab{}.
\newblock \showarticletitle{A neural corpus indexer for document retrieval}.
\newblock \bibinfo{journal}{\emph{Advances in Neural Information Processing Systems}}  \bibinfo{volume}{35} (\bibinfo{year}{2022}), \bibinfo{pages}{25600--25614}.
\newblock


\bibitem[Wu et~al\mbox{.}(2024b)]%
        {dense_2}
\bibfield{author}{\bibinfo{person}{Shiguang Wu}, \bibinfo{person}{Wenda Wei}, \bibinfo{person}{Mengqi Zhang}, \bibinfo{person}{Zhumin Chen}, \bibinfo{person}{Jun Ma}, \bibinfo{person}{Zhaochun Ren}, \bibinfo{person}{Maarten de Rijke}, {and} \bibinfo{person}{Pengjie Ren}.} \bibinfo{year}{2024}\natexlab{b}.
\newblock \showarticletitle{Generative retrieval as multi-vector dense retrieval}. In \bibinfo{booktitle}{\emph{Proceedings of the 47th International ACM SIGIR Conference on Research and Development in Information Retrieval}}. \bibinfo{pages}{1828--1838}.
\newblock


\bibitem[Wu et~al\mbox{.}(2024a)]%
        {hi-gen}
\bibfield{author}{\bibinfo{person}{Yanjing Wu}, \bibinfo{person}{Yinfu Feng}, \bibinfo{person}{Jian Wang}, \bibinfo{person}{Wenji Zhou}, \bibinfo{person}{Yunan Ye}, \bibinfo{person}{Rong Xiao}, {and} \bibinfo{person}{Jun Xiao}.} \bibinfo{year}{2024}\natexlab{a}.
\newblock \showarticletitle{Hi-gen: Generative retrieval for large-scale personalized e-commerce search}.
\newblock \bibinfo{journal}{\emph{arXiv preprint arXiv:2404.15675}} (\bibinfo{year}{2024}).
\newblock


\bibitem[Xiong et~al\mbox{.}(2021)]%
        {ance}
\bibfield{author}{\bibinfo{person}{Lee Xiong}, \bibinfo{person}{Chenyan Xiong}, \bibinfo{person}{Ye Li}, \bibinfo{person}{Kwok{-}Fung Tang}, \bibinfo{person}{Jialin Liu}, \bibinfo{person}{Paul~N. Bennett}, \bibinfo{person}{Junaid Ahmed}, {and} \bibinfo{person}{Arnold Overwijk}.} \bibinfo{year}{2021}\natexlab{}.
\newblock \showarticletitle{Approximate Nearest Neighbor Negative Contrastive Learning for Dense Text Retrieval}. In \bibinfo{booktitle}{\emph{9th International Conference on Learning Representations, {ICLR} 2021, Virtual Event, Austria, May 3-7, 2021}}. \bibinfo{publisher}{OpenReview.net}.
\newblock
\urldef\tempurl%
\url{https://openreview.net/forum?id=zeFrfgyZln}
\showURL{%
\tempurl}


\bibitem[Yuan et~al\mbox{.}(2024)]%
        {gdr}
\bibfield{author}{\bibinfo{person}{Peiwen Yuan}, \bibinfo{person}{Xinglin Wang}, \bibinfo{person}{Shaoxiong Feng}, \bibinfo{person}{Boyuan Pan}, \bibinfo{person}{Yiwei Li}, \bibinfo{person}{Heda Wang}, \bibinfo{person}{Xupeng Miao}, {and} \bibinfo{person}{Kan Li}.} \bibinfo{year}{2024}\natexlab{}.
\newblock \showarticletitle{Generative Dense Retrieval: Memory Can Be a Burden}. In \bibinfo{booktitle}{\emph{Proceedings of the 18th Conference of the European Chapter of the Association for Computational Linguistics (Volume 1: Long Papers)}}. \bibinfo{pages}{2835--2845}.
\newblock


\bibitem[Zhang et~al\mbox{.}(2025a)]%
        {zhang2025hiergr}
\bibfield{author}{\bibinfo{person}{Fuwei Zhang}, \bibinfo{person}{Xiaoyu Liu}, \bibinfo{person}{Xinyu Jia}, \bibinfo{person}{Yingfei Zhang}, \bibinfo{person}{Zenghua Xia}, \bibinfo{person}{Fei Jiang}, \bibinfo{person}{Fuzhen Zhuang}, \bibinfo{person}{Wei Lin}, {and} \bibinfo{person}{Zhao Zhang}.} \bibinfo{year}{2025}\natexlab{a}.
\newblock \showarticletitle{HierGR: Hierarchical Semantic Representation Enhancement for Generative Retrieval in Food Delivery Search}. In \bibinfo{booktitle}{\emph{Proceedings of the 63rd Annual Meeting of the Association for Computational Linguistics (Volume 6: Industry Track)}}. \bibinfo{pages}{444--455}.
\newblock


\bibitem[Zhang et~al\mbox{.}(2025b)]%
        {zhang2025multi}
\bibfield{author}{\bibinfo{person}{Fuwei Zhang}, \bibinfo{person}{Xiaoyu Liu}, \bibinfo{person}{Xinyu Jia}, \bibinfo{person}{Yingfei Zhang}, \bibinfo{person}{Shuai Zhang}, \bibinfo{person}{Xiang Li}, \bibinfo{person}{Fuzhen Zhuang}, \bibinfo{person}{Wei Lin}, {and} \bibinfo{person}{Zhao Zhang}.} \bibinfo{year}{2025}\natexlab{b}.
\newblock \showarticletitle{Multi-level Relevance Document Identifier Learning for Generative Retrieval}. In \bibinfo{booktitle}{\emph{Proceedings of the 63rd Annual Meeting of the Association for Computational Linguistics (Volume 1: Long Papers)}}. \bibinfo{pages}{10066--10080}.
\newblock


\bibitem[Zhang et~al\mbox{.}(2022)]%
        {zhang2022mind}
\bibfield{author}{\bibinfo{person}{Fuwei Zhang}, \bibinfo{person}{Zhao Zhang}, \bibinfo{person}{Xiang Ao}, \bibinfo{person}{Dehong Gao}, \bibinfo{person}{Fuzhen Zhuang}, \bibinfo{person}{Yi Wei}, {and} \bibinfo{person}{Qing He}.} \bibinfo{year}{2022}\natexlab{}.
\newblock \showarticletitle{Mind the gap: Cross-lingual information retrieval with hierarchical knowledge enhancement}. In \bibinfo{booktitle}{\emph{Proceedings of the AAAI conference on artificial intelligence}}, Vol.~\bibinfo{volume}{36}. \bibinfo{pages}{4345--4353}.
\newblock


\bibitem[Zheng et~al\mbox{.}(2024)]%
        {lc-rec}
\bibfield{author}{\bibinfo{person}{Bowen Zheng}, \bibinfo{person}{Yupeng Hou}, \bibinfo{person}{Hongyu Lu}, \bibinfo{person}{Yu Chen}, \bibinfo{person}{Wayne~Xin Zhao}, \bibinfo{person}{Ming Chen}, {and} \bibinfo{person}{Ji-Rong Wen}.} \bibinfo{year}{2024}\natexlab{}.
\newblock \showarticletitle{Adapting large language models by integrating collaborative semantics for recommendation}. In \bibinfo{booktitle}{\emph{2024 IEEE 40th International Conference on Data Engineering (ICDE)}}. IEEE, \bibinfo{pages}{1435--1448}.
\newblock


\bibitem[Zhou et~al\mbox{.}(2022)]%
        {ultron}
\bibfield{author}{\bibinfo{person}{Yujia Zhou}, \bibinfo{person}{Jing Yao}, \bibinfo{person}{Zhicheng Dou}, \bibinfo{person}{Ledell Wu}, \bibinfo{person}{Peitian Zhang}, {and} \bibinfo{person}{Ji-Rong Wen}.} \bibinfo{year}{2022}\natexlab{}.
\newblock \showarticletitle{Ultron: An ultimate retriever on corpus with a model-based indexer}.
\newblock \bibinfo{journal}{\emph{arXiv preprint arXiv:2208.09257}} (\bibinfo{year}{2022}).
\newblock


\bibitem[Zhu et~al\mbox{.}(2024)]%
        {cost}
\bibfield{author}{\bibinfo{person}{Jieming Zhu}, \bibinfo{person}{Mengqun Jin}, \bibinfo{person}{Qijiong Liu}, \bibinfo{person}{Zexuan Qiu}, \bibinfo{person}{Zhenhua Dong}, {and} \bibinfo{person}{Xiu Li}.} \bibinfo{year}{2024}\natexlab{}.
\newblock \showarticletitle{CoST: Contrastive Quantization based Semantic Tokenization for Generative Recommendation}. In \bibinfo{booktitle}{\emph{Proceedings of the 18th ACM Conference on Recommender Systems}}. \bibinfo{pages}{969--974}.
\newblock


\end{thebibliography}


\end{document}